# The Backwards-Time Interpretation of Quantum Mechanics – Revisited With Experiment

Paul J. Werbos, NSF*, pwerbos@nsf.gov
Ludmila Dolmatova, IntControl*, LDOLM@erols.com

**Abstract**

The classic paper of Clauser et al proved that Bell's Theorem experiments rule out all theories of physics which assume locality, time-forwards causality and the existence of an objective real world. The Backwards-Time Interpretation (BTI) tries to recover realism and locality by permitting backwards time causality. BTI should permit dramatic simplification of the assumptions or axioms of physics, but requires new work in fundamental mathematics, such as new tools for the "closure of turbulence," the derivation of statistics generated by ODE or PDE.
      Recent events like the Delayed Choice Quantum Eraser experiment of Kim, Shih et al have increased mainstream interest in the possibility of backwards causality. The Backwards Time Quantum Teleportation (BTQT) experiment will take this further. True backwards time communication channels (BTCC) are absolutely impossible in most formulations of quantum theory but only <u>almost impossible</u> in the BTI formulation. This paper discusses BTQT, the issue of backwards causality in different versions of quantum theory and new mathematical developments. Predictions of a common toy field theory (real $\varphi^3$ QFT) are reproduced by the corresponding PDE model combined with a new model of the micro/macro interface (2M/M).

______________________________________________________________________________________

## 1. Introduction and Guide

The issues raised in the abstract require consideration of many scientific disciplines, ranging from experimental physics through to fundamental mathematics. This paper will summarize our present understanding of these issues, but it will also discuss unresolved questions in all these disciplines, which will require greater attention from specialists in these disciplines. Thus in presenting this material, we will divide it up into three relatively self-contained sections, focused on different audiences and issues:

- Section 2 will focus on BTI, BTQT, causality, and the foundation of physics
- Section 3 will discuss the general mathematical issue of deriving the statistical dynamics of ODE and PDE systems (using tools applicable to SDE as well)
- Section 4 will apply tools related to those of section 3 to derive the cross-time correlations of a real scalar $\varphi^3$ field theory

Section 2 is mainly written for the physicist. Section 3 is mainly written for mathematicians who may have little prior background or interest in quantum theory as such.

      Section 2 will first give some background on BTI. Section 2.2 will discuss the background of the BTQT experiment now in process and its general characteristics. In BTQT, each bit of information from a chip will be downloaded at a "transmission time" t and "received" at a time equal to t minus a microsecond. However, BTQT is *not* capable of true backwards information communication; as with ordinary quantum teleportation, the quantum message cannot be decoded until a classical control bit is also received. Thus the significance of the experiment is mainly theoretical, not technological.

      Section 2.3 will review why true backwards time communication channels (BTCC) are absolutely impossible, according to conventional versions of quantum field theory (QFT). Section 2.4 will analyze the general challenge of how to modify QFT -- either in mainstream or neoclassical forms – so as to accommodate backwards causality. It will derive a simple model (2M/M) of the interface between the

microscopic realm, in which dynamics and causality operate in a time-symmetric manner, and the macroscopic realm, which is dominated by the forwards arrow of time in thermodynamics. Because of the equivalence between PDE+2M/M and conventional QFT, for many ordinary field theories, this theory does not permit the construction of Backwards Time Communication Channels (BTCC). However, while 2M/M is superior to the simple model 1M/M assumed both in classical physics and in classical QFT, it is still only an approximation, in principle; thus there is some possibility, in theory, that a true BTCC might be possible someday – but the options we could imagine are so remote that most physicists would not take them very seriously. Section 2.6 will discuss the general requirements of trying to reconstruct QFT using a combination of PDE, statistical mathematics and 2M/M; prior to that, section 2.5 will provide some general qualitative arguments to try to address the concerns of those physicists who have been brought up to believe that the use of PDE is a heresy even more severe than the assumption of backwards causality.

Section 3 will shift gears abruptly, to the mathematical side. This section will mainly address a question in pure mathematics which can be studied without regard to possible applications in physics. The question is: how can we map the probability distribution, Pr(state), for the state of a dynamical system, into a new mathematical object or representation which permits more tractable computation of quantities like equilibrium probabilities or equilibrium probabilities conditional upon initial conditions? Strictly speaking, this is only *part* of the mathematics underpinning the physics of section 2.6.2. However, section 3 will also develop certain trace identities for computing expectations values of fields, which underlie the way in which the 2M/M model is expressed in section 2.6.

The main work of this section will begin in section 3.3, where we finally define such concepts as Fock space and specify the new representations of Pr(state). However, we need to do some preliminary work before we reach that point. Section 3.1 will describe the goals of section 3 in more detail, and their relevance to other fields like fluid dynamics, plasma physics, nonlinear system dynamics and biocomplexity. Section 3.2 will discuss prior research on the key questions of this section. It will describe the limitations of existing representations of Pr(state), and explain how we will use those representations as part of the foundation for building something more powerful. (In essence, it is a challenge to overcome a curse of dimensionality when working with the function Pr as such; but the usual representation in terms of statistical moments or correlations leads to a problem with infinite regress or the "closure of turbulence.")

The remainder of section 3 will discuss the dynamics of the new representations in selected examples from first-order ODE, second-order ODE and the nonlinear Klein-Gordon equation. Some of the new representations – especially, the normalized entropy matrix $S_w$ – appear to solve the closure-of-turbulence problem more effectively than quantum theory does. Others – the **z** and $S_z$ representations, based on with second-order systems – obey linear dynamics governed by an antiHermitian gain matrix, exactly the same kind of dynamics seen in conventional quantum field theory (QFT). But these representations all require *two* sets of creation and annihilation operators, in order to handle the general case of arbitrary initial conditions. QFT, by contrast, only requires one set to characterize the nonlinear Klein-Gordon equation. Section 4 will analyze the discrepancy, and discuss a modified approach – formulated as a true four-dimensional approach – which leads to an exact equivalence between the Feynman integrals of a classical $\varphi^3$ theory and those of the corresponding QFT. The larger implications are discussed in section 2.

## 2. The Backwards Time Interpretation: Background, Experiment, Theory

## 2.1. Where we are on BTI

For decades, most physicists and philosophers have been very uncomfortable with formulations of QFT whose predictions depend in principle on questions like "Do cats have souls? I.e. do they qualify as 'observers,' or are they capable of existing in a mixed state?" There has been widespread sympathy for notions like "Quantum Theory Without Observers" [Goldstein] – but very little connection between such efforts and mainstream QFT or mainstream real analysis in mathematics. There have been several reasons for this discrepancy:

- Conventional Wisdom postulates that such issues are purely philosophical. It postulates that they cannot affect our predictions of experiments, let alone the development of new technology. But mathematical developments so far suggest that PDE+2M/M is equivalent to the usual second-quantization QFT (2QFT) only for *some* field theories (probably including the standard model and quasilinear supersymmetry theories), and that it will lend itself to nonperturbative analysis more easily than 2QFT. As we try to reach beyond the perturbative analysis of the standard model, these differences will become ever more important.
- The classic book by Bell [Bell] has performed a great service in popularizing and explaining many aspects of the Bell's Theorem literature, starting from the fundamental theorem originally proven by Clauser et al [Clauser et al]. However, few of its readers seem to have gone back to the original paper by Clauser et al which gives equal weight to *three* assumptions – locality, "hidden variables" (reality) and "causality" (*time-forwards* causality). Thus theories which relax locality have received more attention but BTI remains unknown to most physicists.
- While BTI should permit drastic simplification of the underlying assumptions or axioms of physics – and opens the door to a truly axiomatic mathematical approach – it still requires extensive new mathematical development, Lagrangian by Lagrangian, just as complex as that of 2QFT.
- Above all, the notion of backwards-time causality has been seen as a kind of heresy by physicists imbued with classical ways of thinking and Conventional Wisdom.

Long ago, Von Neumann expressed the view that quantum mechanics appears to violate everyday notions about causality. In the 1960's, DeBeauregard [DeBeauregard] spearheaded the idea that the key paradoxes in quantum theory could be understood by reference to backwards time causality. In that general time frame, Louis DeBroglie [DeBroglie] hinted at the possibility of re-establishing a local realistic theory by chucking out traditional time-forwards notions of causality – but recoiled from it, as something difficult to imagine and formalize in mathematical terms. In 1973, Werbos [Werbos73] pointed out more explicitly that we could recover reality and locality, with reference to the theorem of Clauser et al, by building models which accommodate backwards-time causality. This, in our view, was the birth of the Backwards Time Interpretation.

Nevertheless, it is a long way between a general interpretation and approach, versus the kind of worked out mathematics one sees in standard texts like [Weinberg] or [Zinn-Justin]. Bits and pieces of the required analysis may be found in [Werbos89,93,98,99] – to be updated here. Anthony Leggett has informed us that he too has written extensively on the "forgotten loophole" in the theorem of Clauser et al. Penrose, in the text of his book *Shadows of the Mind* [Penrose], cited the earlier work of DeBeauregard and Werbos, and he has developed viewgraphs illustrating the idea of BTI in Bell's Theorem experiments, viewgraphs which Hameroff has shown at various conferences. But all of this is just a logical starting point, as this paper will show. Considerably more development is needed both on the mathematical and experimental side even to assess the potential benefits of this approach.

## **2.2. The Story Behind the Experiment**

The emerging field of Quantum Information Science (QIS) has begun to force some revision of the Conventional Wisdom about backwards causality and about the importance of quantum foundations. For example, many theorists entering the field have been forced to learn that probability amplitudes as such are not powerful enough to express the state of our knowledge/uncertainty regarding the state of a physical system; formulations of QFT based on density matrices have been essential to the development of working experiments (e.g., [Gershenfeld et al]).

Current-day QIS is mainly based on extensions of the classic Einstein-Podolsky-Rosen (EPR) paradox, which was first translated from theory to experiment in the classic work of Clauser et al [Clauser et al], widely popularized by Bell [Bell]. It builds upon the more recent high-precision tests of "Bell's Theorem" by the Rochester group led by Mandel [Ou et al] and the Maryland Group led by Karol Alley [Alley et al], including Shih. The Department of Defense (DOD) has funded large-scale efforts to implement QIS, and has also supported extensions of the fundamental work at Maryland and Rochester. As part of that work, Kim, Shih and others [Kim,Shih et al] recently performed an experiment called the delayed choice quantum eraser experiment.

The delayed choice quantum eraser experiment was originally inspired by suggestions from Marlan Scully [Scully] and John Wheeler [Wheeler]. Scully proposed a variation of the classic double slit experiment. In Scully's vision, the photons going off to the "right" sometimes interfere with each other, like waves, but sometimes do not, like particles. The choice between wave-like versus particle-like behavior is controlled by events on the "left," following standard QFT. Wheeler suggested modifying the vision such that the key events on the "left" actually occur ***after*** the effects of the choice have been measured "on the right." Shih and Kim found a way to design a concrete experiment to implement these visions of Scully and Wheeler [Kim, Shih et al]. These experiments were successful. They showed that the "choice" of a path by the photon on the "left," encountering a beam splitter, affects the results of a measurement on the right.

This prior experiment was not definitive in proving that causal effects can propagate backwards through time. But it was highly suggestive. Just as fundamental theory needs to be modified, in order to reflect the empirical evidence about the importance of density matrices, this experiment suggests that it should be modified in order to accommodate some limited degree of backwards causality in nature.

In August 1999, Shih and Werbos met each other at the Quantum Computing Program Review in Adelphi, Maryland. (Proceedings are available from Henry Everitt of the Army Research Office, Research Triangle Park, North Carolina.) In his presentation, Shih discussed how current experiments in QIS strongly suggest the need to re-evaluate the role of time in QFT. Werbos expressed his agreement to Shih, during the coffee break, and they exchanged email and telephone numbers. Shih sent Werbos some recent reprints and preprints, and Werbos initiated an email discussion in which he proposed a slight modification of the delayed choice experiment, in order to make the conclusions more visible and graphic.
A meeting was held between Werbos, Dolmatova and Shih in Arlington to discuss the details and initiate the project. At the time, we called this the "Backwards Time Telegraph" experiment, but we soon realized that the expression "Backwards Time Quantum Teleportation" would give a more accurate (albeit still simplified) impression. The funding of this experiment was managed by Gernot Pomrenke, on rotation to NSF from AFOSR.

The exact details of the experiment will be given in the planned future paper by Shih et al, after the results are in. The key ideas are as follows. A message or "code" is stored in a microchip. A part of the code is downloaded into the BTQT proper at time $t_+$. A measurement based on this message is taken at time $t_-$. The time $t_+$ is about a microsecond later than $t_-$. There is no other channel from the code to the measurement, or from the system generating the code to the measurement. (Various methods of code generation will be tried.) The "transmission" is based on use of an optical switch in place of the "BSA" and "BSB" beam splitters in the quantum eraser experiment [Kim, Shih et al], and the "received message" is the same $R(x)$ measurement by the D0 detector of those experiments. Optical fibers are used to provide the delay element to extend the time difference to about one microsecond.

## 2.3. Why BTCC is Impossible in Traditional Versions of QFT

In recent months, there have been many reports about "arguments" whether faster than light (FTL) or backwards-time flows of information are "theoretically possible." The key point is that there are *different theories* here – different formulations of quantum theory – which yield different predictions about what is or is not possible.

Strictly speaking, there are dozens upon dozens of formulations of QFT in the literature, ranging from very formal concepts developed by philosophers through to phenomenological versions developed to cope with various types of experiment. One may classify these formulations based upon how they would address the Bell's Theorem experiments: (1) theories which give up reality; (2) theories which give up locality; (3) theories which give up time-forwards causality; (4) theories which give up more than one. Formulations which do not attempt to explain these core experiments will not be considered here. Based on Occam's Razor, we would propose to focus on the first three types of theory, until and unless there is compelling evidence to tell us that the complexities of the fourth type are unavoidable. (In any case, we are not sufficiently aware of theories of the fourth type to provide a good review of them.)

Theories which give up reality have traditionally been the mainstream formulation of QFT. In this version, QFT is not a theory about "what exists" in an objective, real universe of any sort. It is simply a set of calculating procedures, to predict the results of experiments. All experiments can be ultimately reduced to scattering experiments (or to spectral calculations, which can be posed in terms of scattering). Thus the recipe for predicting scattering experiments is the real foundation here; everything else is simply adornment. (For example, there is no fundamental difference between the Schrodinger picture, the interaction picture and the Heisenberg picture; they all yield the same predictions.)

The traditional recipe for predicting scattering experiments is based on the "second quantization" (2QFT) developed by Feynman, Schwinger and Tomonoga, summarized in the classical text by Mandl [Mandl] and expounded upon in authoritative texts such as Bogolyubov [Bogolyubov], Itzykson and Zuber[IZ], and Weinberg [Weinberg]. The recipe has three stages: (1) all of our knowledge of initial conditions – i.e. of every incoming particle and field, whether elementary or compound – is represented by a wave function $\Psi(t_-)$ over Fock space; (2) a "Schrodinger equation" of the form $\partial_t\Psi=iH\Psi$ is used to calculate $\Psi(t_+)$ for asymptotic future times $t_+$, starting from the initial $\Psi(t_-)$; (3) probabilities of measurement outcomes or "observations" are calculated by associating a measurement operator M to each possible measurement variable, and by computing $\Psi^H M\Psi$, for example, to compute the expected value of such a measurement. When multiple measurements occur in series, it is assumed that the first measurement "condenses the wave function" to an eigenfunction of the corresponding measurement operator.

It is easy enough to update this fundamental recipe by replacing $\Psi$ by $\rho$, the density matrix, a matrix over Fock space, and by eliminating the unnecessary concept of measurements in series. In the quantum foundations literature, the Von Neumann-Wigner interpretation is often invoked to get rid of intermediate measurements – a very important development. (It is also invoked at times to justify the idea of God as the one true Cosmic Observer "necessary" to close physics. We do not mean to criticize religion in general, but this particular piece of theology reminds us of the Dark Ages. Still, it is a major step up from earlier versions of quantum theory in which people may debate whether cats have the metaphysical power to "condense the wave function of the entire universe.") In experimental quantum computing, intermediate measurement devices often involve measurement delay times and noise; thus they are often modeled as physical bodies or as "compound particles" rather than metaphysical observers.

The traditional recipe – either in its original form or updated -- clearly involves a forwards-time flow of information from $t_-$ to $t_+$. Period. There has occasionally been speculation that the Schrodinger equation *over Fock space* might permit FTL or even backwards-time effects related somehow to entanglement; however Bell [Bell] explains very clearly why this is not possible. Eberhard's theorem [Eberhard] demonstrates that FTL communication is impossible in the standard formulations of QFT. If BTCC were possible in such formulations, it would be easy enough to construct a device which violates Eberhard's theorem; therefore, a BTCC is impossible in such formulations.

Eberhard's theorem does *not* explicitly assume time-forwards causality as an axiom. It relies on basic relationships involving equal-time commutators, which are very general in nature. However, it does implicitly assume that the transmitted impulse and the received impulse can be represented somehow as local field operators, like the standard unrenormalized operators $\phi(\underline{x},t)$. This does open the door to a very limited kind of FTL. If the physical particles used to transmit or receive an impulse are themselves not localized, FTL effects are possible when we base our measurements on things like the center of mass of a particle. More precisely, the renormalized field operators used to represent a physical particle are not localized, as assumed by Eberhard, and therefore admit limited FTL effects. This is the basis for Hegerfeldt's explanation of a classical FTL paradox due to Fermi, which is described eloquently on http:/biols.susx.ac.uk/Home/John_Gribbin/quantum.htm. In July 1999, Werbos [Werbos00] pointed out that this phenomenon could be extended in theory by exploiting situations where the renormalized wave function extends over a very long distance. Nevertheless, all of these effects are consistent with 2QFT. They do not violate 2QFT, and they do not contradict those aspects of 2QFT which would make a BTCC impossible.

We have occasionally asked ourselves the following question: could the mechanics of renormalization be used to design a true BTCC as well? If so, it would have to involve a very drastic kind of *multiparticle* renormalization of the entire BTCC apparatus. The "initial time wave function" would have to be defined in a way which includes the future measurement as part of the present wave function, in effect. This would require replacing the present approaches to renormalization by some sort of extremely complex kludge. The modified QFT would assume, in effect, that "the future is caused by the past, so long as we understand that we define the 'past' in a way which includes the future." It would be more plausible to chuck out time altogether as a fundamental concept.

In summary, 2QFT predicts that a BTCC is impossible, unless one stretches and modifies the theory to the point where it is an unrecognizable kludge.

Most serious alternatives to 2QFT include extensive efforts to make sure that they replicate all the predictions of 2QFT exactly, for all possible theories of physics. Therefore they too would not admit any hope of a true BTCC.

That is all that needs to be said about those alternatives, for the specific purposes of this section. However, there are two such alternatives to 2QFT which do cry out for some comment, in any discussion of backwards causality.

First, the modern Functional Integral formulation of Quantum Field Theory (FIQFT) provides a kind of halfway station between 2QFT and BTI, in many ways. We would view FIQFT [Zinn-Justin] as an elaboration of Schwinger's early proposal for "source theory". [Schwinger]. FIQFT uses classical PDE as part of its core formalism far more than 2QFT does. The PDE are generally time-symmetric, and the Green's function calculations are not defined as a byproduct of a process of evolution in forwards time. Yet in his classes, Schwinger constantly invoked the principle of "causality" (time-forwards causality) to close the theory. The equivalence between FIQFT and 2QFT, in cases where 2QFT is capable of making a prediction, is well established. But FIQFT strongly reduces our dependence on perturbation calculations over Fock space; as a result, it has made it possible to perform nonperturbative calculations which have had a decisive effect on progress in nuclear physics [Zinn-Justin] and in superstring physics [Greene] in recent years.

FIQFT almost seems to be a realistic theory in its own right. The calculations may be summarized in a humorous way: "First God rolled the dice over and over again to create a universe – a symmetric, Euclidean, classical four-dimensional universe (i.e. he simulated a local Markhov Random Field [Werbos98]). Then he looked out upon his work and decided that it was not good, so he hit it and spun it around ninety degrees and left the scene. (A Wick rotation)" This is not truly a realistic model, because a Wick rotation applied to the state of a random infinite lattice yields explosive modes which have no physical counterpart. Nevertheless, recent developments in FIQFT [Zinn-Justin] are an impressive start in the right direction.

Second, the Transactional Interpretation (TI) of John Cramer includes lots of discussions about handshakes between the past and future, as in BTI. But Cramer has been emphatic that he invokes *nonlocality* as his way of addressing the Bell's Theorem results. He has stated that his formalism would agree precisely with 2QFT in all circumstances.

## **2.4. Rebuilding QFT Without Time-Forwards Causality: Fundamental Challenges**

The main goal of this paper is to specify one possible strategy for reformulating QFT, so as to account for backwards causality and restore both reality and locality. This section will try to discuss the options for allowing backwards causality in a more general way – but even here it is easier to start out from one concrete possibility.

Concretely – to a build a theory based on locality and reality, the obvious starting point is to go back to classical ideas about continuous fields defined over Minkowski space-time, governed by a Lagrangian which results in a set of PDE. Again, we will discuss other, more complex options later on; this is just a starting point. (Also, the reader should recall that this sort of classical theory, in the spirit of Einstein and DeBroglie [DeBroglie], is very different from the older sorts of classical theory based on Poisson brackets and h=0 and so on.)

How could we calculate the scattering probabilities predicted by such a "classical" model, for situations like a Bell's Theorem experiment? In the traditional approach, we would control certain variables on the input side, which we may call "**q**". We would assume that the state S of the system at an initial time $t_-$ is governed by some probability $Pr(S(t_-))=p(\mathbf{q}, S(t_-))$. For a Bell's Theorem experiment, we may assume that we have a set of polarizers on the outgoing channels, which are instantly turned on and encountered at a later time $t_+$. Let us denote the settings of these polarizers as $\mathbf{q}_+$. We also have a set of counters which detect the presence or absence (**c**) of photons at a slightly later time $t_{++}$.

In the traditional approach to scattering, we follow a recipe very similar to 2QFT. We encode our initial knowledge into the probability distribution $Pr(S(t_-))$. We then evolve the dynamics, based on the classical PDE, so as to compute $Pr(S(t_+))$. For each possible state $S(t_+)$, we use our knowledge of $\mathbf{q}_+$ to evolve the field state still further forwards in time, to compute **c** for that state. (We are avoiding any discussion of wave-like behavior versus particle-like behavior in this section; see section 2.4 for such details.)

The Bell's Theorem experiments have clearly ruled out any theory which calculates scattering probabilities in this way, for dynamics based on PDE or any other local dynamics. However, this kind of calculating procedure does **not** properly reflect the content of the original PDE!

In the realm of QED, at least, the underlying PDE are symmetric with respect to time. Therefore, as Einstein emphasized over and over again, we need to pull ourselves away from our everyday intuitive attachment to time as a kind of magical "special" dimension. We need to look down upon the entire space-time continuum as a whole, and formulate the mathematics from that viewpoint. This is not an easy thing to do, but it is the primary task which faces us.

From this viewpoint, the PDE ask us to consider the probability of alternative possible states of the fields **across space-time**. We are looking for solutions of the PDE across space-time, subject to a **pair** of boundary conditions, $\mathbf{q}_-$ and $\mathbf{q}_+$, which have equal status. The mathematical challenge is to calculate probabilities of possible states of the fields across all of space-time, subject to the combination of boundary conditions. There is absolutely no reason to assume apriori that $Pr(\mathbf{c} | \mathbf{q}_-, \mathbf{q}_+) = Pr(\mathbf{c} | \mathbf{q}_-)$. That apriori assumption, used in traditional calculations, is based on a common-sense belief that causality can only operate forwards in time; however, that apriori belief cannot be justified on the basis of the PDE themselves, when the PDE are in fact symmetric with respect to time. **This leads to the fundamental mathematical challenge: how *can* we calculate $Pr(c | \mathbf{q}_-, \mathbf{q}_+)$, without making such an apriori assumption, based solely on the PDE themselves?**

There also exists a more traditional, "mainstream" approach to accounting for backwards-time causality. We can proceed by trying to modify a mainstream formulation of QFT, such as the many-worlds theories of Everett and Wheeler or of the Bohmian physicist David Deutsch (who is often regarded as father of quantum computing). In that case, the state of the cosmos $S(t_-)$ at time $t_-$ is described as a wave function over Fock space, rather than a set of fields over three dimensional space. But in all other respects the situation is exactly as we described in the previous paragraphs!!! When we ditch the apriori assumption that $Pr(S(t_+)|\mathbf{q}_-, \mathbf{q}_+) = Pr(S(t_+)|\mathbf{q}_-)$, we then need to find a way to calculate $Pr(S(t_+)|\mathbf{q}_-, \mathbf{q}_+)$ and $Pr(\mathbf{c}|\mathbf{q}_-, \mathbf{q}_+)$ based solely on the assumed dynamical equation, solved in a way which gives equal treatment to the boundary conditions at $t_-$ and at $t_+$. This appears easier to do in the PDE case than in the Everett-Wheeler case; thus even if you believe the Everett-Wheeler version more, it makes sense to try to develop the mathematics in the easier case first.

In either case, we immediately run into some very fundamental difficulties. The underlying dynamics (PDE or other) are in fact symmetric with respect to time. Thus the statistics which they generate all by themselves must be symmetric with respect to time. Yet we cannot really set up an experimental apparatus to run a Bell's Theorem experiment in reverse time, as neatly and as easily as we can in forwards time. We cannot set up a laser which sends photons back through time as easily as we can send them forwards in time.

More formally, the problem here is that $Pr(S|\mathbf{q}_-, \mathbf{q}_+)$ is essentially indeterminate. The boundary conditions $\mathbf{q}_-$ and $\mathbf{q}_+$ plus the PDE do not contain sufficient information by themselves in order to specify this probability distribution. In the traditional approach, people dealt with this situation by assuming that **thermodynamics** provides the missing information. We assume that we have a kind of global probability distribution, $Pr(S(t_-))=P(S(t_-))$ already available. The boundary condition $\mathbf{q}_-$ operates as a kind of filter or constraint; we calculate $Pr(S(t_-)|\mathbf{q}_-)$ simply by throwing out all states S which do not fit $\mathbf{q}_-$, and then by scaling the probabilities P(S) of the remaining states so as to make the probabilities add up to one. This calculating strategy may be called the "first approximate model of the micro/macro interface" (1M/M). It is the basis of the traditional classical approach to scattering discussed – and ruled out – above.

In [Werbos93], Werbos proposed a kind of second generation approximate model of the micro/macro interface, which we may call 2M/M. The idea was to calculate $Pr(S(t_-)|\mathbf{q}_-, \mathbf{q}_+)$ by starting from $P(S(t_-))$, as before, but throwing out all states S which do not fit **both** $\mathbf{q}_-$ and $\mathbf{q}_+$, and then scaling up the probabilities again to add up to one. In order to implement this idea in calculations, one can go through the following stages: (1) calculate a representation of $Pr(S(t_-)|\mathbf{q}_-)$, exactly as one does this in 1M/M;

(2) evolve that representation forwards in time, as in 1M/M, so as to calculate a representation of $Pr(S(t_+)|\mathbf{q}_-)$; (3) <u>then</u> throw out all the states $S(t_+)$ which do not fit $\mathbf{q}_+$, and scale up the probabilities. Under certain conditions (see sections 3 and 4), we can represent $Pr(S(t_-)|\mathbf{q}_-)$ by a positive definite matrix $\rho$ of unit trace defined over Fock space. In that case, for an appropriate representation, step 3 simply reproduces the procedure of calculating $Tr(M_i\rho)$ in order to calculate the probability of observing the i-th possible value of a measured quantity M. In other words, it exactly reproduces the usual measurement formalism of 2QFT. This is sufficient to explain the Bell's Theorem results and the success of 2QFT under ordinary conditions.

There is reason to believe that PDE models of the underlying laws of physics, combined in this way with 2M/M, could provide a more compact axiomatic basis for deriving the predictions of QFT, for a wide variety of possible field theories. However, we have demonstrated this so far only for the case of a simple toy field theory, the real scalar $\varphi^3$ model, described in textbooks like [Brown] and [ZJ]. Sections 3 and 4 describe some of the mathematical challenges involved in taking this further. The underlying assumptions in this approach are far simpler than those of traditional QFT, but there is every reason to believe that a full analysis of complex field theories will still require complex calculations, no matter how simple the underlying assumptions. There is reason to believe that this will probably reproduce the predictions of QFT for quasilinear models with cubic interactions (like the standard model of physics and the usual supersymmetry models). But there is reason to believe that it will *not* be equivalent to 2QFT across all possible models. For example, the combination of PDE and 2M/M will presumably be mathematically well-defined even for complex models such as gravitational models whose "corresponding" models in 2QFT may not be mathematically consistent. This could potentially extend still further the success of nonperturbative methods in opening the door to new models which were not admissible or workable in traditional 2QFT.

Nevertheless, 2M/M is *<u>still only an approximation</u>*. There is need for mathematical work which explores the predictions and potential of PDE+2M/M, but there is also need for work – both theoretical and empirical – which tries to improve upon 2M/M, in representing the interface between macroscopic time-forwards physics and microscopic time-symmetric physics. Even a perfect knowledge of the underlying microscopic laws of physics would not be sufficient for this purpose.

We have explored the possibility of developing a better approximate model, which might be called 3M/M. A few partial, preliminary ideas are mentioned in [Werbos00]. Crudely speaking, we can only imagine four possible mechanisms for allowing a true BTCC:

- A bending of space-time, as imagined in various theories of quantum gravity, which can be analyzed within traditional forms of QFT as well as BTI.
- Mechanisms which would give us more control (more variables $\mathbf{q}_+$) over the time $t_+$ boundary condition, such as "backwards time lasers" that would let us do backwards time Bell experiments. But forwards time lasers are only possible because of our ability to channel sources of forwards-time free energy (like sunlight) through technology to an experiment; to do this in backwards time, we would need either:

  - To discover an external source of backwards-time free energy, like the legendary "black sun" of ancient folklore; or
  - To find a way to create free energy (in the backwards-time direction) out of disorder – the backwards-time equivalent of a perpetual motion machine.

- Justification of a "correction term" to 2M/M, somehow accounting for the fact that disorderly states at time $t_+$ may be inherently more probable, independent of circumstances at time $t_-$. (In other words, based on the theoretical need for a more complete convolution of boundary condition probabilities at $t_-$ and $t_+$.) If this perturbs the commutators at time $t_+$ even slightly, it is conceivable that it might disturb the equal time commutator relations which Eberhard's Theorem is based on, and permit construction of a limited BTCC.

The last of these possibilities may at least serve as a stimulus for trying to develop a better understanding of the micro/macro interface. For example, if this hypothetical correction term should have an exponential (Gibbs-like or even Gaussian) dependence on the ratio between the energy of outgoing photons and the temperature of the corresponding polarizer and counter, could that term be amplified enough to be measurable by manipulating that ratio for one or more outgoing photon? Would there be some possibility of developing something like BTCC or superfast recurrent networks by developing some kind of stochastic infrared quantum computing, based on such effects? [Werbos00]. Michael Conrad has urged us not to dismiss such possibilities too lightly; for example, he argues that some forms of molecular computing might perform very efficient stochastic search, based on something like stochastic infrared quantum computing.[Conrad].

Nevertheless, we do not have more to say about such speculative possibilities at the present time. The remainder of this paper will not discuss such possibilities. Instead, it will focus on the difficult but near-term opportunity to explain as much as we can by combining PDE, statistical mathematics and the 2M/M model.

## 2.5. PDEs as Heresy, or "Are You Folks Joking or What?"

In an ideal world, this paper would proceed immediately to discussing the various mathematical tasks involved in constructing QFTs based on the approach outlined above. However, there are many preconceptions out there about the role of PDEs which may interfere with many readers' ability to focus on the more specific tasks. For example, PDEs are often equated with "classical field theory (CFT)" – though in our view, CFT was really a *combination* of PDEs and the 1M/M model, and the 1M/M model was the source of most of its failings. This section will say a few words about these sorts of preconceptions.

The use of PDE to model the fundamental laws of physics has become something of a heresy to many physicists, particularly to those physicists not aware of the central role of PDEs in modern FIQFT {Zinn-Justin}. There is a strong analogy here to the Conventional Wisdom about neural networks in the 1970's. In 1969, Minsky and Papert [Minsky] argued that artificial neural networks could not be used as a foundation for developing artificial intelligence, or for anything else very useful. The core argument was that multilayer networks are necessary in order to approximate interesting functions, but no one had an algorithm available capable of training such networks. In 1974, Werbos developed such an algorithm, now called "backpropagation," which overcame that argument, in principle [Werbos94, AR]. But it took more than ten years of intense effort, with many twists and turns, before neural networks received any serious attention from the mainstream. The essential problem, during those ten years, was that people had built up all kinds of emotional associations and loose rationalizations for rejecting all possible neural network models. These secondary emotional responses went far beyond the more rational and logical arguments by Minsky. (Unfortunately, some of that debris has yet to be fully cleared away even now.)

The issue of EPR and Bell's Theorem experiments are analogous, in this case, to Minsky's arguments. The backwards time interpretation fully accounts for these arguments, but there is a huge backlog of debris inherited from the Einstein-Bohr debates which would require hundreds of pages to discuss in full. A few heuristic comments may be in order, however:

- Introductory classes in quantum mechanics frequently begin by saying that "classical physics is the limiting case as h goes to zero." They are referring here to Lorentzian point-particle physics and Poisson brackets, not to CFT. Here, we propose to use essentially the same PDE as in FIQFT, where we may adopt the convention that $\hbar=c=1$ but we certainly do not set h=0!
- There are absolutely no "hidden variables" here. The "hidden variables" approach was one early way to try to address the EPR paradox, whose deficiencies are well explained by Bell[Bell]. We are proposing a very different way of addressing that paradox, in which "what you see is what you get," to a greater extent than in conventional QFT.
- It is sometimes said that "in CFD all operators commute." But in order to derive the statistics generated by PDE (see section 3), we find that we need to use the ordinary creation and annihilation operators of QFT, which certainly do not form a totally commutative structure. We do find it easier to generate a bosonic structure than a fermionic structure, as we will discuss, but bosonic QFT is still a fully quantum mechanical structure, and the critical vacuum commutators which underlie the key calculations are certainly not zero.
- In any realistic, local theory, we would ultimately want to address some of the qualitative issues about "waves" versus "particles" discussed in the last century. In our view, DeBroglie's heuristic story about these issues [DeBroglie] is still generally plausible; its major deficiency was the lack of a mathematical apparatus to underpin that story, and to clarify the details. Many of the details will only be truly clear after the mathematical apparatus is more fully developed. Our main difference with DeBroglie is that we view the photoelectric effects as the reflection through time of photon emission, which should not be viewed as a great mystery in any theory which incorporates backwards causality.
- Some physicists believe it would be more plausible to assume a nonlocal real world based on a wave function propagating over Fock space, or a Heisenberg picture equivalent. Crucial to their philosophy is the view that such a dynamical equation, *all by itself*, is sufficient to cause us to see the usual measurement effects of QFT. (For example, see the Ph.D. thesis of Hugh Everett, reprinted in [DeWitt].) Yet section 3 will show that certain statistics of PDE already obey such a dynamical equation. If their philosophy is correct, then this fact *by itself* should be enough to establish the viability of PDE as models of microscopic physics. (In our view, however, additional physical information – as in section 2.4 – is needed to close the system.)

In summary, we believe that the potential power of PDE-based models is well worth reconsidering, even though some physicists regard PDE as an old, discarded idea. Neural networks were regarded as an old discarded idea in the 1970's. String theory was regarded as an old discarded idea at about the same time [Greene]. Even the use of utility functions was regarded as an old, discarded idea for many decades, until Von Neumann and Morgenstern revived it – and other people used it to develop new tools, like the Bellman equation, which is the very foundation of modern learning-based artificial intelligence [Werbos00]. The key question here is as follows: will modern physicists let these kinds of emotional blinders tie their hands yet again, or will they rise to the challenge of finding out what is possible here?

## 2.6. Strategy for Rebuilding QFTs From PDE + 2M/M

Previous sections have discussed many global issues such as how to incorporate backwards causality into a variety of physical models, how to extend the backwards-time interpretation into new physical regimes, etc. This section will focus on a more near-term issue: how could we try to replicate the predictions of a few basic quantum field theories (i.e., 2QFT or FIQFT for a few basic Lagrangians), enough to establish confidence that this approach can accommodate what is presently known from experiments?

### 2.6.1. Strong Versus Weak Replications of QFT: Chaoitons and Fermions

Strictly speaking, there are *two levels* at which we could try to replicate FIQFT or 2QFT – a weak replication or a strong replication. In a weak replication, we can fill in certain gaps by making the exact same set of assumptions which are now used for that purpose in FIQFT. In particular, we can assume:

(Assumption 1) Conditional statistical moments, as in multipoint Green's functions (translated to a density matrix representation), constitute a prediction of the probability of observing outgoing particles at a set of locations and in a set of "spin states."

(Assumption 2) It is legitimate to work with PDE which describe what Schwinger called "classical anticommuting objects" or what Zinn-Justin describes as fields defined over a Grassmann algebra.

Once you have achieved a weak replication of the key predictions of QFT, we would argue that this by itself is enough to establish BTI as being "as good as" FIQFT as a representation of the laws of physics.
      On the other hand, we – like researchers in FIQFT proper – would prefer to see a more complete explanation of the laws of nature. Thus instead of just taking assumption 1 to be an axiom, we would hope to go further, and try to *explain why* certain statistics correspond, in practice, to the probability of seeing a particle. Likewise, before simply invoking the phrase "Grassmann algebras," we would prefer either to explain how such algebras get into physics or, at a minimum, to see a clear, objective specification of the generators of these algebras and their effect on the meaning of the PDE. (Zinn-Justin does briefly state that the generators applicable to any particular scattering experiment are somehow based on the particle sources for that particular experiment; however, we do not immediately see how one would fill in the mathematical

details to make this work across all experiments. If these details have been published somewhere, we apologize for our ignorance of those publications.) We will discuss these issues further in this subsection, but please remember that these are *common problems* for BTI and FIQFT *both*; they are not relevant to the choice between these two ways to formulate quantum theory.

Both in FIQFT and in BTI, the obvious way to try to *explain* assumption 1 is by assuming that elementary particles correspond to something loosely called "solitons" or "instantons [Zinn-Justin, MRS]. In other words, we may assume that they correspond to some sort of "stable" localized pattern in the underlying fields. [Werbos98,99] defines a more precise class of mathematical object, the "chaoiton," which embodies the required properties. Considerably more research will be needed, in order to find out which field theories – whether deterministic or stochastic, topological or simple – give rise to what types of chaoiton. However, it has been established that topologically interesting field theories are very promising [MRS, Greene], and that such field theories can give rise to one type of soliton good enough to fulfill this role. Once we restrict ourselves, *in principle*, to ultimate theories of physics in which the outputs of a scattering experiment will condense into localized outgoing chaoitons, then it follows that the statistical covariance of the fields provides a way of reading out the probability of finding solitons at a desired set of points in space. (The spin states operate in a similar way, except that the 2M/M model must be invoked to establish the completeness of the set of spin states allowed in any given measurement.)

The chaoiton explanation of assumption 1 has some important immediate consequences. Consider, for example, the fact that the usual Maxwell-Dirac system of equations *does not generate chaoitons*. (See [Werbos98,99] for the reasons why, bearing in mind that Maxwell-Dirac is what [MRS] would call a "topologically trivial theory.") Thus Maxwell-Dirac is unacceptable as an ultimate theory of nature, in this view. Both in DeBroglie's final vision [DeBroglie] and in modern superstring theory [Greene], the core elementary particles are assumed to be extremely small bodies, far smaller than anything we can now observe; thus the quasilinear equations which can describe the fields they generate at observable distances are far simpler than the highly nonlinear (or topological) equations which actually hold the core particle together and determine its rest mass. In this view, the ultimate theory of physics should be a "finite theory" (a theory which requires no renormalization), but the gross macroscopic behavior can be approximated well enough by attributing certain infinite adjustments to the point which represents the core of the particle; renormalization is the process which supplies those adjustments.

Within QFT, researchers have been satisfied to classify potential field theories as acceptable (finite or renormalizable) or unacceptable. For a general theory of ODE and PDE statistics, essentially all field theories must be treated as meaningful or acceptable in some sense. The theory must be rooted in more concrete properties, such as the presence or absence of chaoitons, or the ability to *approximate* theories which generate solitons via renormalization, or related sorts of stability properties.

[Werbos99] discusses many issues regarding chaoitons and renormalization where more research is needed; here, however, we will not discuss assumption 1 any further. The remainder of this paper will discuss the challenge of a "weak replication" of QFT, in which we attempt to replicate the same predictions of statistical moments in some form.

Assumption 2 above also leads to some complex issues. In particular, it suggests four possible strategies for replicating QFTs with fermionic elements:

- Follow FIQFT, starting from PDE which contain "classical anticommutative objects." The basis for such approaches would be exactly the same as in FIQFT.
- Establish direct term-by-term equivalence between a QFT containing fermions and a purely bosonic QFT, as was done by [Coleman] for the sine-Gordon model and the Massive Thirring Model in 1+1-D. It is important here that BTI permits a wider range of bosonic models to choose from than the usual forms of QFT.
- Derive fermionic statistics from more rigorous underlying field theories by somehow following the approach of [Hiley] or the approaches of the Russian champions of the Skyrme Model [MRS].
- Assume a fundamental model which is purely bosonic and derive fermions as composite, bound states of bosons, using the special mechanisms described in the classic papers of Goldhaber and of Wilczek, as discussed in [Werbos99].

It seems possible that the second approach might work for replicating QED, without much change in the Maxwell-Dirac equations (in the appropriate representation), but we have not yet had time to check this in the full 4-dimensional approach to be discussed in section 4.

The exploration and evaluation of these strategies is a crucial part of the work ahead, even in a weak replication of standard physics. Nevertheless, the remainder of this paper will focus on the bosonic case, because that is what we have had the time to explore most thoroughly.

### 2.6.2. The Weak Replication of Bosonic 2QFT or FIQFT

Even though 2QFT and FIQFT are mostly equivalent, these two different formalisms look very different. Thus they give us two different targets to try to mach, in replicating conventional field theory.

Let us start out by considering the simplest possible way to try to establish an equivalence with 2QFT. In 2QFT, the wave function $\Psi(t)$ at any time t is a vector over 3-dimensional Fock space. This will be explained more completely in section 3. The wave function is assumed to evolve according to the usual "Schrodinger equation":

$$\partial_t \Psi = iH\Psi, \tag{1}$$

where H is the usual Hamiltonian operator, and the (positive definite Hermitian) density matrix $\rho$ over Fock space evolves according to:

$$\partial_t \rho = iH\rho - i\rho H \tag{2}$$

Given a set of PDE, characterizing the dynamics of some set of fields $\mathbf{j}(\mathbf{x}, t)$ over Minkowski space-time, the obvious way to replicate a 2QFT field theory is as follows. First, find a way to represent the statistics of the field across space $\mathbf{X}$ at any time t. In other words, find a way to map probability distributions $\Pr(\{\mathbf{j}(\mathbf{x},t),$ all $\mathbf{x}\})$ into matrices $\rho(t)$ over Fock space. Then show that the resulting matrix of statistics obeys equation 2 for some Hermitian matrix H. If H happens to correspond to the H of some 2QFT field theory, then this PDE theory might be considered equivalent to that 2QFT field theory, *on these grounds alone*. In any case, it may be considered a kind of valid possible quantum field theory in its own right.

Section 3 will investigate a variety of ODE models and PDE models and a variety of statistical representations, *motivated by the effort* to calculate the statistics generated by those models, *without regard* to the effort to match QFT. Even if PDE+2M/M is equivalent to QFT, based on the use of *different* statistical tools, the tools described in section 3 could nonetheless give us new and valid ways to analyze a wide range of PDE.

If the "many worlds" theories of quantum mechanics were correct, then establishment of equation 2, for the corresponding H, *should be enough by itself* to establish an equivalence between the PDE and the corresponding quantum theory. But if we chuck out those theories of measurement and also give up the Cosmic Cats (as discussed in section 2.3), then all forms of quantum theory have some explaining to do. The 2M/M model (from section 2.4) provides a recipe for providing that explanation; namely:

(1) The initial $\rho(t_-)$ should encode the probability distribution $\Pr(\{\mathbf{j}(t_-)\} \mid \mathbf{q}_-)$.
(2) Equation 2 tells us how to evolve $\rho$ so that $\rho(t_+)$ represents $\Pr(\{\mathbf{j}(t_+)\} \mid \mathbf{q}_-)$.
(3) Measurement probabilities are calculated by calculating $\{ \text{Tr}(M_i\rho(t_+)) \}$ across the set of allowable possible states represented by $M_1,...,M_n$, the orthogonal matrices projecting to the possible density matrix states permitted by the time $t_+$ measurement system.

For reasons discussed in section 2.3, there is no need to allow for "measurement" at intermediate times.

In section 3, we find that a wide range of ODE and PDE can in fact be represented by equation 2, for an Hermitian operator H. Thus they provide a wide range of acceptable possible 2QFT-style field theories to work with.

Unfortunately, the dynamic matrix H which governs the statistics of various PDE appears quite different from the H which results when the same PDE are "quantized" in the usual procedures of 2QFT. This does *not* imply that we cannot find a way to replicate the same quantized field theories; for example, there are two other obvious possibilities: (1) that quantum theories which *appear* different in equation 2 may lead to the same Feynman integrals and scattering probabilities in the end, as in [Coleman];
(2) that we can replicate the same Feynman integrals, in effect, by starting from different PDE, in some cases.

We have looked into this discrepancy more closely for the examples of real scalar $\varphi^3$ and $\varphi^4$ field theories, which are used as examples in many textbooks and in studies of axiomatic QFT. We found that:

- The 3+1-D statistical field theory, as described above, *absolutely required* two sets of creation and annihilation operators, $\{a, a^+, b, b^+\}$, in order to represent the dynamics of the statistics in the general case. But the usual corresponding Schrodinger equation in 2QFT only requires $\{a, a^+\}$. This discrepancy is due to the fact that the task specification in section 3 requires us to be able to start from *any arbitrary* initial probability distribution for time $t_-$.
- If *instead of* representing the initial conditions as an arbitrary probability distribution at time t-, we represent them in terms of "sources" – following the kinds of procedures described by Schwinger, rigorously applicable to PDEs – the computational task is formally equivalent, in some sense, but much easier. This leads to a *four-dimensional* formulation, to be described in section 4, which only requires $\{a, a^+\}$. However, it requires that these operators be defined as operators over *four-dimensional* Fock space, treating time and space on an equivalent footing. For a $\varphi^3$ model field theory, the results appear exactly equivalent to FIQFT, but for $\varphi^4$ they appear different, so far as we can tell.

Because the standard model of physics and the basic forms of supersymmetry theory [Zinn-Justin] are essentially based on cubic interactions rather than quartic interactions, and also because of our preliminary calculations, we are optimistic that a source-based PDE statistical formalism should be able to replicate those models. However, **this has yet to be proven**!

The analysis of section 3 will be far more mathematically grounded than section 4. Thus the challenges for future research include a further development of the tools of section 3, the *extension* of the tools of section 3 to the four-dimensional context of section 4, and the development of a more complete and rigorous form of the material in section 4 – *along with* the application to more realistic field theories! Even though section 4 links more directly to traditional QFT, please bear in mind that the new tools in that section are nonetheless valid for a wide range of dynamical systems. Thus they may lead to new ways to calculate the implications of PDE+2M/M theories, in cases where the more traditional tools break down.

## 3. General Tools for Deriving Statistics of ODE or PDE Systems

## 3.1. Goals and Motivations

This section will describe a new set of tools for addressing some very old questions: starting from the specification of an ODE or PDE system, how can we deduce the probabilities of states of that system, resulting from the dynamics? How can we calculate the evolution of probabilities over time, the equilibrium probability distribution(s), the probabilities for later times conditional upon information from earlier times, and other such quantities of physical interest? One may try to address these questions by explicitly analyzing the behavior of Pr(state), the probability distribution as such; however, it is often easier to answer these questions by mapping Pr(state) into some other type of mathematical object. The goal of this section is to define new such mappings, which improve our ability to answer such questions.
In particular, we will describe a variety of ways to map probability distributions into vectors and matrices over a space called Fock space, to be defined in this section.

Section 2 has shown how these questions may be relevant to quantum foundations. To meet the needs of quantum physics, these questions are important, but *they are not enough*; additional considerations discussed in section 2 and section 4 are also important. But this section will not address those additional considerations. We will try to make this section understandable to mathematicians with no prior background in quantum field theory. We will analyze the representation and dynamics of probability distributions in a manner which physicists would call "classical;" in our view, quantum magic actually results from boundary conditions, like the 2M/M model of section 2, rather than dynamics as such.

Nevertheless, the use of a trace norm to calculate the expectation value of field properties, as used to express the 2M/M model in section 2.6.2, is based on equation 54, as modified to account for reification – all discussed in this section. The matrix $S_z$, normalized to unit trace, plays essentially the same role as the density matrix of quantum field theory.

The questions addressed here are also very important in other areas of science and engineering. For example, in nonlinear system dynamics, the goal is to characterize the qualitative behavior of ODE and PDE and related systems; the questions asked here are an important part of such characterization. In "biocomplexity" or "artificial life," a few researchers are beginning to ask what kinds of dynamical systems result in the complex patterns of correlation across space which we call life. (Traditional thermodynamics often closes its calculations by assuming that the entropy function $S = \log \Pr$ has a particular functional form which prohibits such patterns in a long-term equilibrium; thus to analyze the evolution of such patterns, more general and powerful techniques are necessary, in principle.) In fluid dynamics and in plasma physics, researchers have worried for decades about the problem called "the closure of turbulence" – the problem of how to calculate equilibrium statistical quantities which appear totally intractable using traditional methods. (There are many useful techniques based on "closure assumptions" or approximations, but these impose a certain minimum approximation error in each application; the techniques to be discussed here aim at approximations which, like Taylor series, can converge to the exact correct answer.)

The tools to be described here can be extended fairly easily to the analysis of Stochastic Differential Equations (SDE) [Zinn-Justin], to mixed Forwards-Backwards SDE (FBSDE) [El-Karoui] and to Markhov Random Fields [Werbos98]. However, these extensions will not be discussed here. As with quantum theory, a complete analysis of FBSDE requires the analysis of boundary conditions and such, not just the analysis of dynamics. From a mathematician's point of view, section 2 is just an elaborate explanation of why the microscopic laws of physics may be viewed as the zero-noise limit of an FBSDE to which three spatial dimensions have been appended. We have also considered the possibility of retaining nonzero noise in the model, which would be of interest to many physicists as well; however, the resulting calculations end up being rather involved, and we currently believe that the zero-noise version appears more plausible as a model of the physics we see in current experiments.

The discussion here will be relatively straightforward and inductive. We will try to build up the general picture by proceeding from example to example, without leaving out key details. It should be clear that we are really just wading through the initial shallow waters of a very deep and general mathematical structure. We hope that mathematicians will be interested and able to chart out this structure more completely.

## **3.2. Traditional Versus Fock Space Representations of Probability Distributions**

This section will analyze the dynamics of probability and statistics for two types of dynamical system:

- ODE, in which the state at time t is defined by a vector $\underline{X}(t) \in R^n$.
- PDE over Minkowski space, in which the state at time t is defined by the set of values $\{\underline{j}(\underline{x},t),$ all $\underline{x} \in R^3\}$ for a vector field $\underline{j}$ mapping to $R^n$.

Physicists should note that "$\underline{j}$" at any point is a "vector in $R^n$" as defined by mathematicians; in particular, it may actually be a concatenated combination of scalars, pseudoscalars, covariant vectors, tensors, and so on. Most of this section will deal with the case of $\underline{j}$ which is actually just a scalar (i.e. in $R^1$), but the underlying mathematics is quite general, and we will sometimes make this explicit.

There are two dominant traditional techniques for representing the statistics of ODE or PDE widely used in the English-speaking world. The first is the obvious direct representation: one simply specifies $p(\underline{X})$ or $p(\{\underline{j}(\underline{x},t),$ all $\underline{x}\})$ as some arbitrary function with the required properties (real, nonnegative, adding up to 1), and one studies how this function evolves over time. For example, with an ODE system defined by:

$$\partial_t \underline{X} = \underline{f}(\underline{X}), \tag{3}$$

the function p($\underline{X}$) obeys the well-known Fokker-Planck equation which is usually written as:

$$\partial_t p + \text{div}(p\underline{v}) = 0 \tag{4}$$

which in this case reduces to:

$$\partial_t p + \sum_{i=1}^{n} \frac{\partial}{\partial X_i}(pf_i) = \partial_t p + p\sum_{i=1}^{n}\frac{\partial f_i}{\partial X_i} + \sum_{i=1}^{n}\frac{\partial p}{\partial X_i}f_i = 0, \tag{5}$$

where $f_i$ is the i-th component of the vector-valued function **f**, and where "$\partial_t$" is physicists' notation for the derivative with respect to time. (We would have preferred to just put a dot over the "p," but Microsoft has other preferences.)

Equation 5 does provide a rigorous basis for analyzing the evolution of probabilities. It is free from the problem of "closure of turbulence." Like the Hamilton-Jacobi-Bellman (HJB) equation of control theory and neural networks [Werbos00], this equation is exact and efficient – but it suffers from a severe "curse of dimensionality" when it is used in a direct manner. In control theory, the computational problems with HJB result from the difficulty of representing a value function J($\underline{X}$) (equivalent to the concept of "action" in physics), while here they result from problems in representing a general function p($\underline{X}$). Because of this curse of dimensionality, equation 5 has not been widely used in practice for representing the statistics of highly complex ODE systems (where n is moderate to large) or for PDE in the English-speaking world. (Equation 4 is widely used in other applications which are beyond the scope of this paper, ranging from fluid dynamics to stochastic search.) In neural networks, we overcame the curse of dimensionality by *still using* the fundamental equation, but using it in a different way, with a parametric representation of the function J($\underline{X}$). Likewise, we will argue that equation 5 remains a central tool, but we will propose a different way of using it, based in part on new parametric representations of the function p($\underline{X}$). Representations of p or of log p based on neural networks might be very useful in some applications, like biocomplexity, but this paper will focus on representations based on Fock space.

The second dominant technique is the use of statistical moments. Statistical moments have been the dominant way of representing the statistics of PDE in the English-speaking world. For a PDE system, the moments are usually represented as an infinite sequence of objects $u_0, u_1, .... u_k,...$ defined by:

$$u_0 = 1$$

$$u_1(i, \underline{x}_1) = <\varphi_i(\underline{x}_1)>$$

$$u_2(i, \underline{x}_1; j, \underline{x}_2) = <\varphi_i(\underline{x}_1)\varphi_j(\underline{x}_2)> \tag{6}$$

$$...$$

$$u_k(i_1, \underline{x}_1; i_2, \underline{x}_2;...; i_k, \underline{x}_k) = <\boldsymbol{j}_{i_1}(\underline{x}_1)\boldsymbol{j}_{i_2}(\underline{x}_2)..\boldsymbol{j}_{i_k}(\underline{x}_k)>$$

$$...$$

where the angle brackets denote expectation values and where we have used "i" and "j" instead of the better notation "$i_1$" and "$i_2$" simply in order to minimize printing problems. (Instead of angle brackets, many authors use square brackets, or the letter "E" or the letter "M".) For the ODE case, we use the same definition, except that "$\varphi_i(\underline{x})$" is replaced by $X_i$, and the vectors $\underline{x} \in R^3$ all disappear as arguments.

For many ODE and PDE systems, it is straightforward to derive equations for the dynamics of the moments $\{u_i\}$ which are analogous, in effect, to equation 5. Just as equation 5 is linear in p, these equations will turn out to be linear in the moments. Later parts of this section will give many examples. We will see that the dynamical equations for $\{u_i\}$ look very elegant for those ODE or PDE which are "quasilinear with polynomial interactions;" this includes systems like equation 3, in cases where the functions $f_i$ are polynomial in the variables $\{X_i\}$, and it also includes systems like the standard model of physics (but not general relativity).

Unfortunately, the dynamics of the moments of any order i, $u_i$, generally depend on a combination of $u_i$ *and of* higher-order moments such as $u_{i+1}, u_{i+2}$, etc. This creates a kind of infinite regress problem, in which we cannot calculate the equilibrium value of $u_i$ without explicitly exploiting initial value information for all the moments of higher order, all the way to $u_\infty$ in effect. This infinite regress is the "closure of turbulence problem." (Physicists might imagine a similar problem in quantum field theory, in which the dynamics of $\psi_i$ do depend on higher-order terms. However, in that case, the dynamics of $\psi_i$ *also* depend on lower-order terms, in a symmetrical sort of way – due to the Hermiticity of H – which leads to the well-known tractability in principle of the equilibrium equations.) The usual dynamical equations derived for the moments simply do not have well-defined, unique equilibrium solutions.

Despite this problem, the moments representation can still be very useful, *within the context* of a more complete system. In section 3.3, we will describe how the *collection* of moments $\{u_i\}$ can be considered to be a vector **u** over Fock space. This leads to a certain immediate simplification of the appearance of the dynamics. This simplification is analogous to the use of **y**=A**x** instead of explicitly writing out large numbers of simultaneous linear equations. The simplification does not change the *content* of the equations but, because it is easier to write down, it opens the door to the use of more powerful tools. We will also use the notation "**u**(**X**)" or "**u**(**F**)" to represent the value of **u** for that probability distribution which sets Pr(**X**)=1 or Pr(**F**)=1 for a particular state **X** or **F** of the dynamical system. (Such vectors **u** may be called "pure states." They are not exactly the same as the pure states of physics, but the concept is similar.)

In our view, the problem of the closure of turbulence is essentially a mathematical illusion, which results from the fact that most vectors **u** over Fock space are not *physically realizable*. In other words, they cannot be represented as a convex combination of pure states. They violate the requirements for a set of moments which correspond to a valid probability distribution [GS, Lukacs, PR]. In effect, the usual dynamical equations for **u** *appear* indeterminate because they allow for the possibility of states which are not physically realizable. Unfortunately, it is not so easy to solve this problem by simply projecting **u** to the subspace spanned by physically realizable states. (Perhaps a better mathematician could find an effective way to do this.) We will propose two complementary approaches to overcoming these problems:

- We will define several *alternative* representations of the statistical moments (**y**, **w**, **z**) which are based on transformations of the vector **u**, which yield more tractable dynamics. We will refer to these representations (and to **u** itself) as "primal representations" of the statistics.
- We will define several possible *dual* representations of the probability distribution p and of the entropy function S = log p which eliminate the problem of nonrealizable states (at the price of introducing an issue of uniqueness of representation).

In all cases, the new representations will be defined as vectors or matrices over Fock space; therefore their precise definition must be postponed until after section 3.3, where the basic tools and methods of Fock space will be introduced.

The two dominant traditional representations – raw probability functions and statistical moments $\{u_i\}$ – are certainly not the only representations of probability distributions used in the world. There are at least two other approaches of some importance to the general case here – the approach due to Wiener [Wiener] and the approach based on generating functions or functionals.

Wiener actually claimed that his approach could explain the foundations of quantum theory, in the spirit of what we are doing here. However, he developed his approach before the modern structures of 2QFT and FIQFT were worked out. His formalism has been used in a few places, such as fluid mechanics, but it differs in many ways from the type of structure we are exploring here. Certainly the formalism is valid. Certainly it would be an interesting research project to explore the exact nature of the equivalence between Wiener's formalism and the other formalisms described here. But we will not discuss the Wiener formalism further, because it would be a distraction from the goals of the present paper.

The approach based on generating functions is a very different matter.

The mathematical machinery of generating functions is used very widely in modern physics. Thus in building bridges from mathematics to physics, formalisms based on generating functions are just as promising, in principle, as formalisms based on Fock space.

Generating functions have also been widely used in mathematical statistics. However, in the English speaking world, they seem to be used mainly for analyzing probability functions Pr(X) where X is a scalar [GS]. This applies even to [Lukacs] which, from our limited computer searches, still seems to be the primary source text in English. Lukacs, in his text, emphasizes that the most important developments in this topic required translation from Russian and French into English.

Prokhorov and Rozanov[PR] discuss the case of Pr($\underline{X}$), for $\underline{X} \in R^n$, at some length. Neither we nor our Russian colleagues have been able to locate any effort to apply this kind of representation to the closure of turbulence, even though there are Russian universities where the same students study both that textbook and plasma physics as part of the same degree program.

In the end, even though we have chosen to start our analysis in Fock space, we have also found some straightforward equivalences between Fock space representations and generating function representations. We will discuss these in section 3.3. However, there are many opportunities for mathematicians to deepen our understanding of such equivalences and thereby strengthen the overall structure.

Last but not least, [Zinn-Justin] makes frequent use of generating function methods, both in quantum physics proper and in discussion of topics like SDE. He also relies heavily on the Fokker-Planck equation, in an operator-based style. Even though [Zinn-Justin] is the most thorough and complete analysis we have seen of the mathematical foundations of FIQFT, there are many opportunities to strengthen the mathematical underpinnings still further. Perhaps [Zinn-Justin] is implicitly relying on a much more extensive formal literature in French, in which generating functions and characteristic functions play a larger role. Unfortunately, neither of us has current access to that literature. Dolmatova is grateful that DeBeauregard almost succeeded in setting up a dialogue on these issues between her and the Institut Poincare in 1997; however, this paper would never have been written if she had not agreed to come to Washington at that same time instead.

In summary, there are four traditional formalisms used to analyze the dynamics of Pr(state) in ODE or PDE – direct representation of the function Pr(state), statistical moments $\{u_i\}$, Wiener's formalism and statistical generating functions. The remainder of this section will focus on representations of Pr(state) in terms of a vector or matrix over Fock space, such as the vector $\underline{u}$ which is essentially the same as $\{u_i\}$. We will not discuss Wiener's formalism, but we will show some interesting equivalences to the generating function approach, which is highly developed by statisticians in Russia and in France.

## **3.3. Basic Tools and Definitions of Fock Space For Statistics**

### **3.3.1.  Three Primal Representations of Statistical Moments (u, v and w)**

Let us define $F(R^n)$ and $F(R^3 \rightarrow R^n)$, respectively, as the Fock space defined over $R^n$ (used in representing Pr($\underline{X}$) for $\underline{X} \in R^n$) and the Fock space defined over fields $\underline{j}$ which map from $R^3$ to $R^n$ at any time t. For simplicity, we will use the symbol $\Phi$ to represent the state of a field $\underline{j}$ across all space at time t. We will also use the abbreviation $F(\Phi)$ to represent the corresponding Fock space.

In section 3.2, we stated that the "vector" $\underline{u}$ defined as the collection of the objects $u_i$ defined in equations 6 will be treated as a "vector over Fock space." Implicitly, then, we were defining Fock space as the tensor sum of the various spaces which these objects were defined over. Thus for the ODE case we may define $F(R^n)$ as:
$$R^1 \oplus R^n \oplus (R^n \otimes_s R^n) \oplus (R^n \otimes_s R^n \otimes_s R^n) \oplus \ldots\ldots \text{ to infinity} \qquad (7)$$

where we are using "$\otimes_s$" to denote the symmetrized tensor product. (More concretely, for example, we are assuming $\otimes_s$ such that $R^5 \otimes_s R^5$ would be the 15-dimensional vector space formed by all 5 by 5 real symmetric matrices. Likewise, $u_3(i,j,k)$ is required to be symmetric with respect to *any* permutations of its three inputs.) $F(R^n)$ is thus a true vector space, with infinite but denumerable dimensions. $F(\Phi)$, by contrast, is really the tensor sum of a sequence of *function spaces*, representing functions of spaces of growing numbers of dimensions. Each of the objects $u_i$ is a "vector" in function space, representing a symmetric function of i points in $R^3$.

The $L^2$ norm for a vector $\underline{u}$ in $F(R^n)$ or $F(\Phi)$ is simply the sum, from i = 0 to $\infty$, of the $L^2$ norm of $u_i$. These norms are defined in the usual way; for example, in the ODE case:

$$\left|u_2\right|^2 = \sum_{i,j=1}^{n} \left|u_2(i,j)\right|^2$$

(8)

(Note the absence of a coefficient like (1/n!) or (1/2) in this definition! We have tried to be careful with such coefficients, but in hundreds of pages of calculations it is difficult to be perfect.)

Mathematicians will immediately note that the vector **u** as we have defined it will usually not have a finite, bounded $L^2$ norm!! Mathematicians would have a special interest in Fock-Hilbert space, the subspace of Fock space for which the $L^2$ norm is well-defined and finite. The wave functions $\psi$ of 2QFT are simply vectors of unit norm defined over the complex version of $F(\Phi)$. The density matrices are nonnegative Hermitian matrices of norm $Tr(H)=1$ over the same space.

For *first-order ODE*, defined by equation 3, with $\underline{X} \in R^n$, the same raw moments may be written in a more convenient (but equivalent) form:

$$u_{i_1, i_2, \ldots i_n} = < X_1^{i_1} X_2^{i_2} \ldots X_n^{i_n} >$$

(9)

In this case, we may define the *scaled moments* v as:

$$v_{i_1, i_2, \ldots, i_n} = \frac{u_{i_1, i_2, \ldots, i_n}}{\sqrt{i_1! i_2! \ldots i_n!}}$$

(10)

This implicitly defines a new vector **v** in $F(R^n)$. Note that the $L^2$ norm of this scaled vector will normally be finite. Thus **v** will normally be a vector in Fock-Hilbert space. This vector **v** is a more convenient representation in other ways as well, as we will see.

In this section, we will also explore *second-order* ODE. We will begin by exploring ODE of the general form

$$\partial_t^2(\underline{X}) = \underline{f}(\underline{X}),$$

where **f** is essentially any analytic function. We will then pay special attention to ODE of the form:

$$\partial_t^2(X_i) = -m_i^2 X_i + f_i(\underline{X}), \quad (11)$$

where each function $f_i(\underline{X})$ is a polynomial made up of quadratic or cubic or quartic terms, and where $m_i$ is required to be real. This is not such a narrow special case as it may appear at first. For example, if we allowed **f** to include linear terms, we could simply change coordinates to the eigenvectors of the corresponding matrix. Equation 11 essentially reflects the usual assumption in quantum theory that the linearized PDE should not include explosive or dissipative modes.

To analyze second-order systems, it is important to realize that the state of the system at time t is defined by the *combination* of $\underline{X}(t)$ and $\partial_t \underline{X}(t)$. We will find it most convenient to represent equation 11 as a system of *first-order* ODE in the mutually conjugate vectors $\underline{X}^+$ and $\underline{X}^-$ in $C^n$ defined by:

$$X_k^+ = \frac{1}{2}\left( X_k + \frac{1}{im_k} \dot{X}_k \right)$$

(12a)

$$X_k^- = \frac{1}{2}\left( X_k - \frac{1}{im_k} \dot{X}_k \right)$$

(12b)

In this case, it is convenient to write the raw statistical moments as:

$$u_{i_1,\ldots,i_n; j_1,\ldots,j_n} = < (X_1^+)^{i_1} (X_2^+)^{i_2} \ldots (X_n^+)^{i_n} (X_1^-)^{j_1} \ldots (X_n^-)^{j_n} >$$

(13)

In this case, it is most convenient to defined the scaled moments as:

$$v_{i_1,...,i_n;j_1,...,j_n} = u_{i_1,...,i_n;j_1,...,j_n} \sqrt{\frac{m_1^{i_1+j_1} m_2^{i_2+j_2} ... m_n^{i_n+j_n}}{i_1!...i_n! j_1!...j_n!}}$$

(14)

This definition of **v** will also have a finite norm, under ordinary conditions, but it will lead to a simpler expression for the statistical dynamics which result from equation 11.

Finally, this section will also consider PDE defined by the classic nonlinear Klein-Gordon equation, which physicists typically write as:

$$(\Box + M^2)\varphi = f(\varphi) ,$$ (15)

which may also be written as:

$$\partial_t^2 \varphi = \Delta\varphi - M^2\varphi + f(\varphi)$$ (16)

where $\Delta$ is the usual Laplace operator defined by:

$$\Delta \boldsymbol{j} = \sum_{i=1}^{3} \frac{\partial^2 \boldsymbol{j}}{\partial x_i^2}$$

(17)

If we perform a Fourier analysis of $\varphi(\underline{x})$ in equation 16, equation 16 becomes:

$$\partial_t^2 \varphi(\mathbf{p}) = -|\mathbf{p}|^2 \varphi(\mathbf{p}) - M^2\varphi(\mathbf{p}) + (\iint ... \varphi\varphi...)$$ (18)

where we use the physicists' convention of writing the Fourier transform of $\varphi(\underline{x})$ as "$\varphi(\mathbf{p})$" and where the fuzzy term in parentheses indicates some sort of convolution integral which we cannot specify more exactly until we chose a specific function $f(\varphi)$. Equation 18 may be viewed as a modified version of equation 11, in which the index "i" is replaced by the vector "**p**". We may define the scaled vector **v** to be defined by the corresponding modification of equation 14, with $m_i$ replaced by sqrt($|\mathbf{p}|^2+M^2$). This yields the definition:

$$v_n(\underline{p}_1^+,...,\underline{p}_n^+;\underline{p}_1^-,...,\underline{p}_m^-) = u_n(\underline{p}_1^+,...,\underline{p}_n^+;\underline{p}_1^-,...,\underline{p}_m^-) \sqrt{\frac{w(\underline{p}_1^+)w(\underline{p}_2^+)...w(\underline{p}_n^+)w(\underline{p}_1^-)...w(\underline{p}_m^-)}{n_1^+!n_2^+!...n_a^+!n_1^-!...n_b^-!}}$$

(19)

where $\omega(\mathbf{p})$ is defined as sqrt($|\mathbf{p}|^2+M^2$), where $\alpha$ is the number of *distinct* values for **p** in the sequence $\mathbf{p}_1^+,...\mathbf{p}_n^+$, where $n_i^+$ represents the number of occurrences in that sequence of the i-th distinct value, and where $\beta$ and $n_i^+$ are likewise for the sequence $\mathbf{p}_1^-,...,\mathbf{p}_m^-$. Notice that there are n+m quantities in the numerator of the fraction, but only $\alpha+\beta$ in the denominator.

Equation 19 is somewhat disquieting to the mathematician, because it depends on the number of <u>exact duplicates</u> that a vector **p** may have in the list of arguments. It is similar to other situations in physics where the Dirac delta function, a discontinuous distribution, plays a very important role. Perhaps there is a more elegant way to deal with the situation; however, the apparent singularity here is not just an artifact of crude mathematics. "Multiple occupancy" states, in which the exact same value of **p** appears multiple times in the argument list, are the basis of "coherent states" such as "coherent light." That is what a laser is all about.

Finally, there are reasons to define a *normalized* version of **v**, a representation of the statistical moments which have first been scaled and then normalized to a norm of one. This trivial sounding change in representation is far more difficult and far-reaching than it sounds. For example, we will see that **u** and **v** both obey linear dynamical equations; in order to make this true for **w** as well, we cannot just divide the vector **y** by its length. We must exploit the concepts of *pure state* and *physically realizable state* defined in section 3.2.

In the notation of section 3.2, our definitions for the vector **y** (in the ODE case) may be written equivalently as:

$$\underline{\mathbf{y}} = <\underline{\mathbf{v}}(\underline{\mathbf{X}})> = \int \underline{\mathbf{v}}(\underline{\mathbf{X}}) \Pr(\underline{\mathbf{X}}) d^n \underline{\mathbf{X}} \qquad (20)$$

We may define **w**(**X**) for a *pure state* as:

$$\underline{w}(\underline{X}) = \frac{\underline{v}(\underline{X})}{|\underline{v}(\underline{X})|} \qquad (21)$$

For a first-order ODE, defined by equation 3, we may calculate the norm of **y** as defined in equation 10 by simply squaring and summing, which simply yields:

$$|\underline{v}(\underline{X})|^2 = e^{|\underline{X}|^2} \qquad (22)$$

which, inserted into equation (21), gives us:

$$\underline{w}(\underline{X}) = e^{-\frac{1}{2}|\underline{X}|^2} \underline{v}(\underline{X}) \qquad (23)$$

and:

$$\underline{w} = <\underline{w}(\underline{X})> = \int e^{-\frac{1}{2}|\underline{X}|^2} \underline{v}(\underline{X}) \Pr(\underline{X}) d^n \underline{X} \qquad (24)$$

For second-order ODE defined by equation 11, we may calculate the norm of **y** as given in equation 14:

$$|\underline{v}(\underline{X})|^2 = \exp\left(2\sum m_k \left|X_k^+\right|^2\right) = \exp\left(2\sum m_k X_k^+ X_k^-\right) \qquad (25)$$

which likewise leads to:

$$\underline{w} = <\underline{w}(\underline{X})> = \int \exp\left(-\sum m_k X_k^+ X_k^-\right) \underline{v}(\underline{X}) \Pr(\underline{X}) d^n \underline{X} \qquad (26)$$

Finally, for the nonlinear Klein Gordon equation (15), we may take the norm of **y** as defined in equation 19 to get:

$$|\underline{v}(\Phi)|^2 = \exp\left(2\int |\boldsymbol{j}^+(\underline{p})|^2 \boldsymbol{w}(\underline{p}) d^3 \underline{p}\right) \qquad (27)$$

and:

$$\underline{w} = <\underline{w}(\Phi)> = \int \exp\left(-\int \boldsymbol{j}^+(\underline{p}) \boldsymbol{j}^-(\underline{p}) \boldsymbol{w}(\underline{p}) d^3 \underline{p}\right) \underline{v}(\Phi) \Pr(\Phi) d^\infty \Phi, \qquad (28)$$

where equation 28 is a functional integral, a very familiar calculation in physics [Zinn-Justin].

As we will see, all of these vectors **w** in Fock space obey linear dynamical equations. For applications such as the calculation of scattering statistics in physics, for example, it is not unreasonable to restrict our attention possible field states Φ for which the integral in equation 27 is finite.

### 3.3.2. Creation and Annihilation Operators, Clean Versus Physical, and |0>

Creation and annihilation operators are the fundamental building blocks of statistical dynamics in Fock space. They are also the fundamental building blocks of traditional quantum field theory (2QFT). Basic texts like [Mandl] typically specify the operator H in equations 1 and 2 by multiplying and adding various "field operators," like "φ(**x**)", which they then define in turn by combining creation and annihilation operators. From a mathematician's point of view, there is some value in defining H more explicitly and directly in terms of the creation and annihilation operators.

In this section we will define two different types of creation and annihilation operators. First we will define "clean" operators $\{c_i, c_i^+\}$, using a definition which may seem natural and straightforward to the mathematician. Then we will define "physical" operators $\{a_i, a_i^+\}$ based on the definition which physicists normally use. In principle, we could get by, by using only one set of definitions, but we find it interesting to see how the physicists' definition can simplify the mathematics, even in a context where no quantum magic is involved. In calculations with second-order systems, we will need to use a second set of clean operators $\{d_i, d_i^+\}$ and a second set of physical operators $\{b_i, b_i^+\}$.

In the ODE case, for $\underline{X} \in R^n$, the clean operators consist of the operators $c_i$ and $c_i^+$ for i = 1 to n. They are linear operators, which map a vector in Fock space to a vector in Fock space. They have a very simple intuitive interpretation. We may visualize a vector in Fock space as a collection of objects arranged in a ladder – a cell by itself, a line of n cells over that, an n-by-n square of cells over the line, a cube, and so on. Each cell holds just one real number (and the numbers must obey the symmetry requirement we discussed earlier) The creation operator $c_i^+$ moves every number up the ladder, by adding the index "i" to its list of coordinates. For example, the number in cell number k of the line gets moved to the (i,k) cell in the square (and to the equivalent cell (k,i)). The number in the (j,k) cell in the square gets moved to the (i,j,k) cell in the cube. In general, we could say that $c_i^+$ moves the number in cell A to cell (A,i).

The annihilation operator $c_i$ is just the transpose of $c_i^+$. For every cell A, it moves the number in (A,i) down the ladder into cell A. Notice that the numbers in cells which do not have the index "i" in their list of coordinates do not get moved anywhere; they are overwritten and lost. Thus $c_i$ is not an invertible operator.

Creation and annihilation operators are sometimes called "ladder operators."
More formally, if **v** is any vector in $F(R^n)$, we may define:

$$(c_k^+ \underline{v})_{i_1, i_2, \ldots, i_k, \ldots i_n} = v_{i_1, i_2, \ldots, i_k - 1, \ldots i_n}$$
(29)

$$(c_k \underline{v})_{i_1, i_2, \ldots, i_k, \ldots i_n} = v_{i_1, i_2, \ldots, i_k + 1, \ldots i_n}$$
(30)

with the understanding that the right hand side of equation 29 refers to zero whenever $i_k$ is zero. In the special case where n=1, we may visualize **v** as a kind of ladder from bottom to top, such that equation 29 reduces to a matrix multiplication like:

$$\begin{bmatrix} 0 & 1 & 0 & 0 \\ 0 & 0 & 1 & 0 \\ 0 & 0 & 0 & 1 \\ 0 & 0 & 0 & 0 \end{bmatrix} \begin{bmatrix} \ldots \\ v_2 \\ v_1 \\ v_0 \end{bmatrix} = \begin{bmatrix} \ldots \\ v_1 \\ v_0 \\ 0 \end{bmatrix}$$
(31)

Note that $c^+$ is an upper triangular matrix in this picture. Likewise c is just the transpose of $c^+$,

and is lower triangular. By matrix multiplication it is easy to see that $cc^+=I$, but $c^+c$ equals I *minus* the "1" which should appear in the lower right cell of the matrix.

We will also have frequent need to use the number operator $N_k$ defined by:

$$(N_k \underline{v})_{i_1,i_2,\ldots,i_k,\ldots i_n} = i_k v_{i_1,i_2,\ldots,i_k,\ldots i_n} \tag{32}$$

In the case where n=1, the number operator N represents the matrix:

$$\begin{bmatrix} \ldots & \ldots & \ldots & \ldots & \ldots \\ \ldots & 3 & 0 & 0 & 0 \\ \ldots & 0 & 2 & 0 & 0 \\ \ldots & 0 & 0 & 1 & 0 \\ \ldots & 0 & 0 & 0 & 0 \end{bmatrix}$$

By matrix multiplication (or use of the formal definitions) it is easy to recover some important basic identities:

$$c_i^+ = c_i^T \tag{33a}$$

$$N_i^T = N_i \tag{33b}$$

$$(c_i N_i)^T = N_i^T c_i^T = N_i c_i^+ = ((N_i+1)c_i)^T = c_i^T(n_i+1) = c_i^+(N_i+1) \tag{33c}$$

The creation and annihilation operators used in physics differ from the clean operators in only one respect: they include a kind of scaling factor, which simplifies the appearance of the dynamics when we work with the $\underline{v}$ or $\underline{w}$ representations. In the case where n=1, the operator a may be pictured as the matrix:

$$\begin{bmatrix} \ldots & \ldots & \ldots & \ldots & \ldots \\ \ldots & 0 & 0 & 0 & 0 \\ \ldots & \sqrt{3} & 0 & 0 & 0 \\ \ldots & 0 & \sqrt{2} & 0 & 0 \\ \ldots & 0 & 0 & 1 & 0 \end{bmatrix}$$

Also we define $a_i^+ = a_i^T$. More formally, we may define:

$$a_i = c_i N_i^{1/2} = (N_i+1)^{1/2} c_i \tag{34}$$

(Because $N_i$ is a nonnegative diagonal matrix, in effect, $N_i^{1/2}$ is simply the same matrix where each element is replaced by its positive square root.) Transposing this, we also get:

$$a_i^+ = N_i^{1/2} c_i^+ = c_i^+ (N_i+1)^{1/2} \tag{35}$$

and we may deduce:

$$c_i = a_i N_i^{-1/2} = (N_i+1)^{-1/2} a_i \tag{36}$$

$$c_i^+ = N_i^{-1/2} a_i^+ = a_i^+ (N_i+1)^{-1/2} \tag{37}$$

It is also straightforward to verify three identities used very often in physics:

$$a_i^+ a_i = N_i \tag{38}$$

$$a_i a_i^+ = N_i + 1 \tag{39}$$

$$[a_i^+, a_i] \equiv a_i^+ a_i - a_i a_i^+ = -1 \tag{40}$$

where the square brackets are standard notation for the commutator
[A,B] defined as AB-BA. We will make frequent use of the well-known identities:

$$[A, BC] = [A, B]C + B[A, C] \tag{41a}$$

$$[AB, C] = A[B, C] + [A, C]B \tag{41b}$$

By use of mathematical induction and equations 40 and 41, we easily deduce two more useful identities:

$$[(a_i^+)^n, a] = -n(a_i^+)^{n-1} \tag{42a}$$

$$[(a_i)^n, a_i^+] = n(a_i)^{n-1} \tag{42b}$$

Another fundamental tool in Fock space is the concept of the vacuum state |0>. For our purposes here, |0> is simply that vector in Fock space for which $v_0=1$ and all other components $v_i$ of **y** are totally zero. In physics, there is sometimes reason to discuss other more complex concepts of vacuum, but they are beyond the scope of this section.

### 3.3.3. Additional Properties of Realizable u, v, w, and definitions of **r**$_u$, **r**$_v$, **r**$_w$

In section 3.2, we stressed that we are *not* really interested in all vectors **y** in Fock space here. We are interested in vectors which are physically realizable, vectors which represent a mixture of pure states such as **u**(**X**) or **v**(**X**). Before we can exploit the special properties of physically realizable states, we must first begin to analyze what these properties are.

In the simplest case, the case of ODE with $X \in R^1$, the definition in equation 6 reduces to:

$$u_i(X) = X^i \tag{43}$$

For this **u**(X), the definition of c in section 3.3.2 tells us that:

$$(cu)_i = X^{i+1} = X u_i \tag{44}$$

This logic generalizes easily to other ODE and PDE cases:

$$c_i \underline{u}(\underline{X}) = X_i \underline{u}(\underline{X}) \tag{45}$$

$$c_i(\underline{x}) \underline{u}(\Phi) = \varphi_i(\underline{x}) \underline{u}(\Phi) \tag{46}$$

(The physicist may note an uncanny resemblance between equation 46 – an equation which we have *derived* – and definitions of field operators which are commonly *assumed* in quantum theory. The resemblance becomes much stronger when we move from this **u** representation to the **z** representation of a later section. But again, we are not introducing new assumptions here.)
In fact, since these annihilation operators all commute with each other, we may deduce:

$$c_1^{i_1} c_2^{i_2} ... c_n^{i_n} \underline{u}(\underline{X}) = X_1^{i_1} X_2^{i_2} ... X_n^{i_n} \underline{u}(\underline{X})$$

$$\tag{47}$$

and the equivalent relation for the PDE case.

Likewise for $X \in R^1$, equation 10 tells us that:

$$v_i(X) = \frac{X^i}{\sqrt{i!}}$$

(48)

For this vector $\underline{v}(X)$, the definition of a in section 3.3.2 tells us that:

$$(a\underline{v})_i = \sqrt{i+1}\left(\frac{X^{i+1}}{\sqrt{(i+1)!}}\right) = X\left(\frac{X^i}{\sqrt{i!}}\right) = Xv_i$$

(49)

Again, we may easily generalize to all first-order ODE cases:

$$a_i\, \underline{v}(\underline{X}) = X_i\, \underline{v}(\underline{X})$$

(50)

Likewise, since $\underline{w}(\underline{X})$ is just a scalar multiple of $\underline{v}(\underline{X})$, equation 50 implies:

$$a_i\, \underline{w}(\underline{X}) = X_i\, \underline{w}(\underline{X})$$

(51)

in those cases. Because $\underline{w}(\underline{X})$ has unit norm, we may go on to deduce that:

$$\underline{w}(\underline{X})^H a_i\, \underline{w}(\underline{X}) = X_i\, \underline{w}(\underline{X})^H \underline{w}(\underline{X}) = X_i$$

(52)

where the subscript "H" refers to the Hermitian conjugate. (Here, the Hermitian conjugate is the same as the transpose, but most of these calculations carry over easily to the complex case.) We might be tempted to try to calculate the expected value of $X_i$ across a mixture of states simply by calculating the expectation value of equation 52; however, this does not work since $<\underline{w}(\underline{X})^H \underline{w}(\underline{X})>$ does not equal $<\underline{w}(\underline{X})^H><\underline{w}(\underline{X})>$ in the general case!

In order to recover some additional special properties, then, it is convenient to define a new set of representations of the underlying statistics, based on some *matrices* over Fock space. We may define:

$$\rho_u = <\rho_u(\underline{X})> = <\underline{u}(\underline{X})\, \underline{u}(\underline{X})^H>$$

(53a)

$$\rho_v = <\rho_v(\underline{X})> = <\underline{v}(\underline{X})\, \underline{v}(\underline{X})^H>$$

(53b)

$$\rho_w = <\rho_w(\underline{X})> = <\underline{w}(\underline{X})\, \underline{w}(\underline{X})^H>$$

(53c)

Based on these definitions, the expectation value of equation 52 leads immediately to:

$$<X_i> = \text{Tr}(a_i \rho_w)$$

(54)

Again, because the annihilation operators all commute with each other, we can compute expectation values of any polynomial in $\underline{X}$ in the same way.

In order to analyze second-order ODE systems, we will use two sets of operators – $\{a_i^+, a_i\}$ to create and annihilate $X_i^+$ and $\{b_i^+, b_i\}$ to create and annihilate $X_i^-$. In the simplest case, where n=1, equation 14 can be written:

$$v_{i;j}(X^+, X^-) = (X^+)^i (X^-)^j \sqrt{\frac{m^{i+j}}{i!\, j!}}$$

(55)

The definition of the operator a gives us:

$$(a\underline{v})_{i;j} = \sqrt{i+1}\left((X^+)^{i+1}(X^-)^j \sqrt{\frac{m^{i+j+1}}{(i+1)!\,j!}}\right) = X^+ m^{1/2}\left((X^+)^i(X^-)^j \sqrt{\frac{m^{i+j}}{i!\,j!}}\right) = X^+ m^{1/2} v_{i;j}$$

(56a)

For the case of n>1, the calculations are the same for any operator $a_i$, except that the equations are cluttered up by indices and products which are not affected by $a_i$; we still obtain the more general result:

$$a_i \, \underline{y} \;=\; X_i^+ m_i^{1/2} \, \underline{y} \tag{56b}$$

Following essentially the same logic which led from equation 51 to equation 54, we may further deduce:

$$<X_i^+> \;=\; \mathrm{Tr}(\,(a_i/m_i^{1/2})\,\rho_w\,) \tag{57a}$$

$$<X_i^-> \;=\; \mathrm{Tr}((b_i/m_i^{1/2})\,\rho_w) \tag{57b}$$

In the Klein-Gordon case, the Fourier components take the place of variables $X_i$, and we arrive at:

$$<\varphi^+(\mathbf{p})> \;=\; \mathrm{Tr}\left(\,(1/w^{1/2}(\mathbf{p}))\,a(\mathbf{p})\,\rho_w\,\right) \tag{57a}$$

$$<\varphi^-(\mathbf{p})> \;=\; \mathrm{Tr}\left(\,(1/w^{1/2}(\mathbf{p}))\,b(\mathbf{p})\,\rho_w\,\right) \tag{57b}$$

The physicist will note that the operator $(1/w^{1/2}(\mathbf{p}))\,a(\mathbf{p})$ is somewhat similar to the usual field operator used to "quantize" $\varphi(\mathbf{p})$. The correspondence will become exact in section 3.3.6 after we discuss reification. In essence, we have *derived* -- not assumed – the notion of performing measurement by taking the trace of a measurement operator convoluted with a density matrix. As in the ODE case, we can calculate expectation values of any polynomials in $\{\varphi(\mathbf{p})\}$ by inserting the corresponding polynomial of measurement operators into the trace calculation.

Similar calculations allow some interpretation of the creation operators, for physically realizable states. In the case of first-order ODE, for example, one may easily deduce that:

$$\frac{\partial}{\partial X_i} \underline{u}(\underline{X}) = N_i c_i^+ \underline{u}(\underline{X})$$

(58)

and:

$$\frac{\partial}{\partial X_i} \underline{v}(\underline{X}) = a_i^+ \underline{v}(\underline{X})$$

(59)

Differentiating equation 23 and exploiting equation 59, we deduce:

$$\frac{\partial}{\partial X_i} \underline{w}(\underline{X}) = (a_i^+ - a_i)\underline{w}(\underline{X})$$

(60)

We are not entirely sure of how to exploit these relations. By taking the transpose of equation 54, you can see that $\mathrm{Tr}(a_i^+\rho_w)$ is the same as $\mathrm{Tr}(a_i\rho_w)$ here. Perhaps these relations make it possible to derive the Fock space equivalents of classical ODE and PDE more directly than the brute force methods we will use in section 3.4 and thereafter.

Finally, by using the tools of creation and annihilation operators, we can devise alternative expressions for $\underline{v}(\underline{X})$ and $\underline{w}(\underline{X})$ which may of great use in the future, both in making contact with quantum mechanics and in focusing our attention to the case of physically realizable states. In the case of first-order ODE, it is straightforward to show that equation 10 is equivalent to the definition:

$$\underline{v}(\underline{X}) = \exp(\sum_{i=1}^{n} X_i a_i^+) \,|0>$$

(61)

Likewise equation 14 for second-order ODE is equivalent to:

$$\underline{v}(\underline{X}) = \exp(\sum_{i=1}^{n} \sqrt{m_i}(X_i^+ a_i^+ + X_i^- b_i^+)) \,|0>$$

(62)

and equation 19 for the Klein-Gordon equation is equivalent to:

$$\underline{v}(\Phi) = \exp\left(\int \sqrt{\underline{w}(\underline{p})}(\underline{j}^+(\underline{p})a^+(\underline{p}) + \underline{j}^-(\underline{p})b^+(\underline{p}))d^3\underline{p}\right) \,|0>$$

(63)

Equations 23 and 61 easily lead to an expression for $\underline{w}(\underline{X})$ for first-order ODE:

$$\underline{w}(\underline{X}) = \exp(-\tfrac{1}{2}|\underline{X}|^2 + \sum_{i=1}^{n} X_i a_i^+) \,|0>$$

(64)

Likewise, it is easy to combine equation 26 with equation 62, or equation 28 with 63, in order to derive an expression for $\underline{w}(\underline{X})$ or $\underline{w}(\Phi)$ for the second-order dynamical systems.

### 3.3.4 Dual Representations: Entropy Matrix Closes Turbulence

We have theoretically considered 20 different dual representations of Pr(state) at various times, all of them theoretically valid. The most useful appears to be the *entropy matrix*, $S_w$. $S_w$ appears to solve the problem of the closure of turbulence. It does *not* obey the dynamic equations of QFT; instead, it obeys simpler-looking dynamics which seem to offer a kind of direct, feedforward calculation of equilibrium statistics.

$S_w$ is used to represent $Pr(\underline{X})$ or $Pr(\Phi)$ as follows:

$$Pr(\underline{X}) = (1/Z) \exp(-\tfrac{1}{2} \underline{w}(\underline{X})^H S_w \underline{w}(\underline{X})),$$

(65)

where Z is a scalar used to make sure that the integral of $Pr(\underline{X})$ over all $\underline{X}$ will equal one.

In order to explain the significance of $S_w$, we will first discuss a few simpler dual representations.

The simplest dual representation is basically just a Taylor series representation of $Pr(\underline{X})$ or $Pr(\Phi)$. The coefficients of a Taylor series form a vector in Fock space. For example, in the case of an ODE with $\underline{X} \in R^1$, our previous definitions simply give:

$$u_i(X) = X^i$$

(66)

Thus the Taylor series representation of Pr(X),

$$Pr(X) = \Sigma b_i X^i = \Sigma b_i u_i(X) ,$$

(67)

could be written equivalently as:

$$Pr(X) = \underline{b}_u^H \underline{u}(X),$$

(68)

where $\underline{b}_u$ is the vector in $F(R^1)$ made up of the components $\{b_i\}$. Likewise, if $\underline{X} \in R^n$, a Taylor series in the n variables $X_1,...,X_n$ can be written as:

$$Pr(\underline{X}) = \underline{b}_u^H \underline{u}(\underline{X}),$$

(69)

where $\underline{b}_u \in F(R^n)$. It is easy to construct variations of this representation, by replacing "u" in equation 69 by "v" or "w" or "z." (The primal $\underline{z}$ representation will be discussed in section 3.3.6.)

The vector $\underline{b}_u$ defined in equation 69 does obey reasonable-looking dynamics. We will discuss it further. But it has several serious limitations. As with the $\underline{u}$ representation, there is no guarantee that any vector $\underline{b}_u$ in Fock space actually represents a valid probability distribution. When we calculate an equilibrium value for $\underline{b}_u$, there is every possibility that it might represent a function "Pr($\underline{X}$)" which is negative in some parts of space or unbounded in others. The same problem applies to $\underline{b}_v$, $\underline{b}_w$ and $\underline{b}_z$. It would even apply to matrix representations like $Pr(\underline{X}) = \underline{u}(\underline{X})^H B_u \underline{u}(\underline{X})$.

In the entropy matrix representation, we will require that $S_w$ be Hermitian and nonnegative. This is a relatively easy requirement to meet and to verify. It is precisely the same condition that density matrices in QFT are required to meet (though they must also meet the requirement that $Tr(\rho)=1$).

*Any* matrix $S_w$ which meets that requirement will guarantee that the exponent in equation 65 is everywhere nonpositive. This guarantees that the exponential term will have a uniform upper bound of 1. Likewise, the functional form in equation 65 guarantees that $Pr(\underline{X})$ will always be nonnegative. Thus under reasonable conditions, every matrix $S_w$ meeting our basic conditions will imply a unique valid probability distribution. Conversely, the logarithm of a well-behaved probability distribution should always be representable as a Taylor series; we would speculate that a positive definite Taylor series can always be represented in this kind of quadratic form. The arguments in this paragraph cry out for more precise theorems and proofs, but the bottom line seems reasonably clear.

As an alternative to $S_w$, we define $S_u$, $S_v$ and $S_z$ by substituting the letters "u," "v" or "z" for "w" wherever "w" appears in equation 65. We will see that the primal representations $\underline{u}$, $\underline{v}$ and $\underline{w}$ (but not $\underline{z}$!) all obey dynamics like equation 1, but with matrices H which are *not* Hermitian; they are governed by gain matrices which are lower triangular, like the annihilation operators of section 3.3.2. This is the source of the infinite regress problem discussed in section 3.2. We will see that $S_u$, $S_v$ and $S_w$ are governed by a gain matrix which is completely *upper triangular*; this makes it possible to proceed step-by-step, first calculating the first order terms (assuming no knowledge of higher order terms), then the second-order terms, and so on. Clearly $S_u$ and $S_v$ share some of the same advantages as $S_w$. $S_z$, by contrast, is governed by an Hermitian matrix H (i.e., an antiHermitian gain matrix iH), exactly like $\rho$ in quantum theory.

These dual representations do have one major disadvantage – nonuniqueness. More precisely: two different nonnegative Hermitian matrices, $S_w$ and $S_w'$, can correspond to the same probability distribution $Pr(\underline{X})$. This issue will be discussed in detail in section 3.3.7.

The 20 dual representations mentioned above consist of five representations dual to the four representations $\underline{u}$, $\underline{v}$, $\underline{w}$ and $\underline{z}$. For $\underline{u}$, for example, the obvious possible dual representations are $\underline{b}_u$ and $B_u$, which represent a Taylor series to specify Pr(state), while there can also be a vector representation $\underline{\mathbf{b}}_u$ and a matrix representation $S_u$ which represent Taylor series to specify the entropy $S = \log Pr(state)$. A fifth dual representation of interest might be $S_u/Tr(S_u)$.

### 3.3.5. Correspondence Between Fock Space Representations and Generating Function Representations

This paper uses only Fock space representations, in representing and deriving the dynamics of Pr(state). But we would speculate that the entire structure of alternative options and dynamics for Fock space representations is isomorphic to a corresponding structure of options and dynamics for representations based on generating functions. A better understanding of this isomorphism could be useful in strengthening both types of representation, just as dualities in string theory have been very useful in that domain [Greene].

Perhaps the simplest type of generating function is a moments-generating function where, for example, we can calculate the m-th order moments of a random vector $\underline{X} \in R^n$ by calculating the m-th order derivatives of a generating function $G(\underline{s})$:

$$\left. \frac{\partial}{\partial s_{i_1}} \frac{\partial}{\partial s_{i_2}} ... \frac{\partial}{\partial s_{i_m}} \right|_{\underline{s}=0} G(\underline{s}) \;=\; <X_{i_1} X_{i_2} ... X_{i_n}>$$

(70)

In essence, we could simply define G(**s**) as a Taylor series in **s**. We could define it as that Taylor series whose coefficients are just the moments vector **u**(**X**), to within some scalar factors! But those Taylor series coefficients are best described as a vector in the Fock space F($R^n$). More generally, any generating function G(**s**) may be considered as a Taylor series in **s**; the Taylor series coefficients of that function correspond to a vector in Fock space. Likewise, for any vector **y** in Fock space it is possible to define a corresponding generating function something like

$$G(\underline{s}) = \underline{y}^H \underline{u}(\underline{s}) \tag{71a}$$

or

$$G(\underline{s}) = \underline{y}^H \underline{w}(\underline{s}) \tag{71b}$$

At first sight, the Fock space formulation seems somewhat richer, especially since we make heavy use of *matrices* over Fock space. But those matrices could be considered as the Taylor series coefficients for a symmetric generating function G($\underline{s}_1$, $\underline{s}_2$).

Some of the identities in section 3.3.3 should seem familiar in spirit to statisticians experienced in working with generating functions.

Generating functions have some advantages in forcing mathematical rigor, and in drawing on a large literature (in Russian and in French) in mathematical statistics. Fock space representations appear more explicit and intuitive in some ways, and easier to tie in with quantum theory. Perhaps a better understanding of the underlying equivalence would make it possible to combine the advantages of both.

### 3.3.6 Reification and z: The Bridge to Quantum Theory

Section 3.3.1 defined three primal representations of Pr(state) – the raw moments vector **u**, a modified version of **u** scaled in order to have a bounded $L^2$ norm (**y**), and a modified version of **y** scaled in effect to unit norm (**w**). We will see that all three of these representations obey lower triangular dynamics. All three are very different from the dual representations, like $S_u$, $S_v$ and $S_w$, which obey purely upper triangular dynamics. The relation between **w** and $S_w$ is analogous in part to the relation between a vector and a dual vector, when both are expressed in a nonorthogonal coordinate system. ***Reification*** is a further transformation of **y** or **w**, which is analogous to the orthogonalization of a coordinate system. It eliminates the asymmetry between primal and dual representations. When successful, it results in dynamics like that of equations 1 or 2, in which H is an Hermitian matrix.

We have found it useful to define four auxiliary operators for use in reification. For the case of second-order ODE over **X** in $R^n$, these are:

$$M_+(c) = \exp\left( c \sum_{i=1}^{n} a_i^+ b_i^+ \right) \tag{72}$$

$$M_-(c) = \exp\left( \overline{c} \sum_{i=1}^{n} a_i a_i \right) \tag{73}$$

$$S_+(c) = \exp\left( c \sum_{i=1}^{n} \left( a_i^+ a_i^+ + b_i^+ b_i^+ \right) \right) \tag{74}$$

$$S_-(c) = \exp\left( \overline{c} \sum_{i=1}^{n} \left( a_i a_i + b_i b_i \right) \right) \tag{75}$$

We have mainly studied the use of reification for second-order systems, where there is a duality between operators $a_i$ and $b_i$, representing complex conjugate variables. However, in deriving the dynamics of **w** for first-order ODE, we will explain how the basic machinery works. We have found that reification can be applied surprisingly easily to the examples of first-order ODE considered here. For second-order PDE systems, we will replace "$a_i^+ b_i^+$", for example, by $a^+(\mathbf{p})b^+(-\mathbf{p})$, because complex conjugation has the effect of reversing the sign of $\mathbf{p}$ in $\exp(i\ \mathbf{p}^H\mathbf{x})$. This particular form of reification only applies to systems which obey CPT symmetry. The letter "M" refers to "Mixing," while "S" refers to "self." In earlier notes, we found that $M_+(c)$ and $M_-(c)$ seem to form a symplectic algebra, but we have not followed up on that analysis.

In working with second-order systems, we have actually found two different forms of reification which result in antiHermitian gain matrices (as in QFT). For systems represented in terms of conjugate vectors, $\underline{\mathbf{X}}^+$ and $\underline{\mathbf{X}}^-$, we have defined:

$$\underline{\mathbf{z}} = M_-(-1)M_+(-½)\ \underline{\mathbf{y}} \qquad (76)$$

For second-order systems represented by the original real variables $X_i$ and $(\partial_t X_i)/m_i$, created and annihilated by $\{a_i^+, a_i\}$ and $\{b_i^+, b_i\}$, respectively, we have defined:

$$\underline{\mathbf{z}} = S_-(-½)S_+(-¼)\ \underline{\mathbf{y}} \qquad (77)$$

As we were typing this paper, we realized that this same transformation easily reifies the real first-order ODE which we have been working with as well.

We are also intrigued by the fact that equation 24 could be written equivalently (for physically realizable states) as:

$$\underline{\mathbf{w}} = S_-(-½)\ \underline{\mathbf{y}}, \qquad (78)$$

where the terms involving "$b_i$" are deleted from the definition of $S_-$ (because they are not applicable in first-order ODE). We suspect that these definitions are the tip of a very large iceberg – a deep mathematical structure.

As we were typing this paper, another side of this iceberg came into view. Combining equations 78 and 61, we may derive:

$$\underline{\mathbf{w}}(\underline{\mathbf{X}}) = S_-(-½)\exp(\Sigma X_i a_i^+)\ |0\rangle = (\ S_-(-½)\exp(\Sigma X_i a_i^+)(S_-(-½))^{-1}\ )\ (S_-(-½)\ |0\rangle) \qquad (79)$$

The two big terms on the right-hand-side can be simplified very easily. The leftmost term is just the same similarity transform to be discussed in section 3.4.4. In the rightmost term, if we express $S_-(-½)$ as a Taylor series, we will see that all but one of the terms involves some product of annihilation operators, all of which result in zero when applied to the vacuum state; thus $S_-(-½)|0\rangle$ is simply $|0\rangle$. Using the methods of section 3.4.4 to perform the similarity transform, we end up with another expression or definition for $\underline{\mathbf{w}}$ for the case of first-order real ODE:

$$\underline{\mathbf{w}}(\underline{\mathbf{X}}) = \exp(\Sigma X_i(a_i^+ - a_i))\ |0\rangle\ \ !!! \qquad (80)$$

With second-order PDE, the same logic would introduce factors like $w(\mathbf{p})$. It begins to look very, very similar to what is used in calculations involving vacuum probability amplitudes in quantum field theory.

Reification cannot be applied successfully to *all* dynamical systems. After all, it would be very surprising if closed dissipative systems which converge to a point equilibrium or "heat death" could be represented somehow as systems which conserve a proper norm!! In section 3.4.12, we will find that reification based on equation 77 is successful only for those first-order systems which are "statistically noncompressive" – a new concept which we will introduce in section 3.4.6. In section 3.5, we will see that *all* the second-order ODE considered here are statistically noncompressive. Reification has also worked on the nonlinear Klein-Gordon equation. We have only done a few preliminary calculations for other PDE suggested by quantum field theory, but it seems reasonable for now to hope that those theories which have empirical evidence behind them can be reified. Methods like the $S_w$ matrix analysis, which is *not* so similar to present-day quantum theory, may be useful in analyzing those theories which cannot be reified.

**3.3.7 Equivalence Classes for Real $S_w$ and $r_w$, and Null Matrices**

Section 3.3.4 mentioned that two different matrices, $S_w$ and $S_w$', can represent the same probability distribution $Pr(\underline{X})$ or $Pr(\Phi)$, even if both are nonnegative and symmetric (Hermitian). In that case, we will say that $S_w$ and $S_w$' are *equivalent*. Our definition of $S_w$ in equation 65 tells us that we may express this notion of equivalence more formally (for the case of real-valued ODE systems) as:

$$( S_w \equiv S_w' ) \Leftrightarrow (\forall \underline{X})(\underline{w}(\underline{X})^T S_w \underline{w}(\underline{X}) = \underline{w}(\underline{X})^T S_w' \underline{w}(\underline{X})) \tag{81}$$

If we define the symmetric (Hermitian) matrix $\delta$ as $S_w - S_w'$, then equation 81 tells us that $\delta$ is a *null operator*, where we define:

$$(\delta \text{ null}) \Leftrightarrow (\forall \underline{X})(\underline{w}(\underline{X})^T \delta \underline{w}(\underline{X}) = 0) \tag{82}$$

It is easy to see that the equivalence classes for $S_w$ form nice parallel linear subspaces of $F(R^n) \otimes F(R^n)$.

We believe that there is a straightforward algorithm for determining whether any operator D is null. Suppose that D may be written as $D(a_1, a_1^+, a_2, a_2^+, ... a_n, a_n^+)$, where D is a polynomial or Taylor series in these creation and annihilation operators. Then, exploiting the algebra of operator commutators, it is possible to write:

$$D = \sum_{\boldsymbol{a}} D_{\boldsymbol{a},1}(a_1, a_1^+) D_{\boldsymbol{a},2}(a_2, a_2^+)...D_{\boldsymbol{a},n}(a_n, a_n^+) \tag{83}$$

where D has been expanded into a sum (finite or infinite) of terms indexed by $\alpha$, and where each of the terms $D_{\alpha,i}$ is itself a sum of the form:

$$D_{\boldsymbol{a},i} = \sum_{k,\ell=0}^{\infty} (D_{\boldsymbol{a},i})_{k,\ell} (a_i^+)^k (a_i)^\ell \tag{84}$$

This is more or less the same as the "normal form" used in calculations in quantum field theory[Mandl].

For any such operator D (whether null or not), we may define the *reduced* version of D, D', as:

$$D' = \sum_{\boldsymbol{a}} D'_{\boldsymbol{a},1}(a_1) D'_{\boldsymbol{a},2}(a_2)...D'_{\boldsymbol{a},n}(a_n) \quad , \tag{85}$$

where:

$$D'_{\boldsymbol{a},i} = \sum_{k,\ell=0}^{\infty} (D_{\boldsymbol{a},i})_{k,\ell} (a_i)^{k+\ell} \tag{86}$$

For any *pure state*, $\underline{w}(\underline{X})$, equation 51 and its transpose tell us that:

$$\underline{w}(X)^T D \underline{w}(X) = \underline{w}(X)^T \left( \sum_{\boldsymbol{a}} \prod_i \sum_{k,\ell} (D_{\boldsymbol{a},i})_{k,\ell} (a_i^+)^k (a_i)^\ell \right) \underline{w}(X)$$

$$= \sum_{\boldsymbol{a}} \prod_i \sum_{k,\ell} (D_{\boldsymbol{a},i})_{k,\ell} \underline{w}(X)^T (a_i^+)^k (a_i)^\ell \underline{w}(X) = \sum_{\boldsymbol{a}} \prod_i \sum_{k,\ell} (D_{\boldsymbol{a},i})_{k,\ell} \underline{w}(X)^T (X_i)^k (X_i)^\ell \underline{w}(X)$$

$$= \sum_{\boldsymbol{a}} \prod_i \sum_{k,\ell} (D_{\boldsymbol{a},i})_{k,\ell} \underline{w}(X)^T (X_i)^{k+\ell} \underline{w}(X) = \sum_{\boldsymbol{a}} \prod_i \sum_{k,\ell} (D_{\boldsymbol{a},i})_{k,\ell} \underline{w}(X)^T (a_i)^{k+\ell} \underline{w}(X)$$

$$= \underline{w}(X)^T D' \underline{w}(X) \tag{87}$$

Based on this calculation, we may deduce that D is null if and only if D' is null. But because D' is made up entirely of destruction operators, it can be null over the full range of $\underline{\mathbf{w}}(\underline{\mathbf{X}})$ only if it is identically zero. Thus we can test for the nullness of D by reducing D to D', and finding out whether the result is identically zero or not.

Going through the same calculations for $D^T$, it is easy to see that $(D^T)'=D'$. Thus $D^T$ is null if and only if D is null. Furthermore, $(D+D^T)'=2D'$, which tells us that D is null if and only if the symmetrized version of D is null. It is easy to find examples of symmetric null operators, such as $a^+a^+-2a^+a+aa$.

Equation 82 may be turned on its head to address the issue of determining whether a matrix $\rho_w$ is physically realizable (as $\rho_w$). For any physically realizable matrix $\rho_w$, equation 53c tells us that:

$$\text{Tr}(D\rho_w) = <\text{Tr}(D\rho_w(\underline{\mathbf{X}}))> = <\text{Tr}(D\ \underline{\mathbf{w}}(\underline{\mathbf{X}})\ \underline{\mathbf{w}}(\underline{\mathbf{X}})^T)> = <\underline{\mathbf{w}}(\underline{\mathbf{X}})^T D\underline{\mathbf{w}}(\underline{\mathbf{X}})> = <0> = 0$$

(88)

for all null matrices D. We go on to conjecture that $\rho_w$ is physically realizable if and only if $\text{Tr}(D\rho_w)=0$ for all symmetric null matrices D.

Next let us recall that the space of all real symmetric matrices $\rho_w$ is $F(R^n)\otimes_S F(R^n)$ – the same as the space which the matrices $S_w$ are taken from. The equivalence classes for $S_w$ provide a partitioning of this space. The exact same partitioning of the space $F(R^n)\otimes_S F(R^n)$ is also useful in analyzing matrices $\rho_w$! For each *physically realizable* matrix $\rho_w$, we may define the set of *related matrices* $\rho_w'$ as the set of matrices $\rho_w+D$, where D is a symmetric null matrix. The set of $\rho_w$ and its related matrices is exactly the same space as one of the equivalence classes for $S_w$! Starting from any matrix $\rho_w'$, we may find the related physically realizable matrix $\rho_w$, simply by a kind of Gram-Schmidt process, subtracting out multiples of null matrices D based on the inner product $\text{Tr}(D\rho_w')$. Notice how critical it is to treat matrices as if they were vectors here; this general approach may be useful in other aspects of the future development of this area.

This last observation suggests that we can go back to the case of matrices $S_w$ and define one more useful concept. We may define the *canonical form* of $S_w$, representing any probability distribution $\text{Pr}(\underline{\mathbf{X}})$, as the one version of $S_w$ within the equivalence class which has the property that $\text{Tr}(S_w D)=0$ for all symmetric null matrices D. This concept has not played a major role in our work so far, but we can imagine ways in which it could become useful in the future. Likewise, we have not yet had time to explore the variations of these concepts to the cases of $S_z$ and $\rho_z$, to second-order systems, and so on. (For the complex case, $D_{\alpha,i}$ in equations 83 and 84 would be a function of $a_i$, $b_i$, $a_i^+$, and $b_i^+$, where $a_i$ and $b_i$ multiply $\underline{\mathbf{w}}(\underline{\mathbf{X}})$ by $X_i$ and its complex conjugate, respectively. In that case, we would use Hermitian conjugates instead of matrix transposes, but the resulting structure and conclusions would be very similar.)

## 3.4 Dynamics of Representations of Pr(X) for First-Order ODE

### 3.4.1 Goals, Cases of ODE To Be Studied, and Definition of Gain Matrix G

This section will derive the dynamics of most of the representations defined above, in some typical examples of first-order ODE. It will focus almost exclusively on the case where $\underline{\mathbf{X}}\in R^n$ and where the function $\mathbf{f}$ in equation 3 is given by:

$$f_i(\underline{X}) = \sum_{j=1}^n A_{ij} X_j + \sum_{j,k=1}^n B_{ijk} X_j X_k$$

(89)

In order to avoid confusing notation, it will mainly demonstrate the details for the case n=3, and extrapolate the results to other n. The actual details of the calculations should make it obvious that this is a valid extrapolation. In the case where n=3, the notation for the components of $\underline{\mathbf{u}}$ and $\underline{\mathbf{y}}$ in equations 9 and 10 leads to straightforward-looking equations.

Throughout this paper, whenever a vector $\underline{\mathbf{y}}$ in Fock space obeys linear dynamics, we will

write this as:
$$\partial_t \underline{y} = G_v \underline{y}, \tag{90}$$

and we will refer to $G_v$ as the "gain matrix" of $\underline{y}$. Likewise, the Hermitian matrices $\rho$ over Fock space discussed here will obey linear dynamics governed by:
$$\partial_t \rho = G_\rho \rho + \rho G_\rho^H, \tag{91}$$

where $G_\rho$ is the gain matrix of $\rho$.

The first goal of this section will be to derive the gain matrices for most of the representations defined above, for dynamical systems defined by equations 3 and 89. Other goals are to explain the larger significance of these representations, and to use these examples to illustrate principles of duality and the machinery of reification.

### 3.4.2. Dynamics of u for ODE Given In Eqs. (3) and (89)

For the case where n=3, equation 9 gives:
$$u_{ijk} = < X_1^i X_2^j X_3^k > \tag{92}$$

Let us first work out the dynamics for the case where $B_{ijk}$ in equation 89 is all zero. From basic calculus, we deduce:
$$\partial_t(X_1^\ell X_2^m X_3^n) = \ell(X_1^{\ell-1} X_2^m X_3^n)(\partial_t X_1) + m(X_1^\ell X_2^{m-1} X_3^n)(\partial_t X_2) + n(X_1^\ell X_2^m X_3^{n-1})(\partial_t X_3) \tag{93}$$

Substituting in from equations 3 and 89 (again, without "$B_{ijk}$") and taking expectation values on both sides, we may deduce:
$$\partial_t u_{\ell mn} = \ell A_{11} u_{\ell mn} + \ell A_{12} u_{\ell-1,m+1,n} + \ell A_{13} u_{\ell-1,m,n+1}$$
$$+ m A_{21} u_{\ell+1,m-1,n} + m A_{22} u_{\ell mn} + m A_{23} u_{\ell,m-1,n+1}$$
$$+ n A_{31} u_{\ell+1,m,n-1} + n A_{32} u_{\ell,m+1,n-1} + n A_{33} u_{lmn} \tag{94}$$

If we use the definitions of "clean" operators from section 3.3.2, equation 94 can be expressed more elegantly as:
$$\underline{\dot{u}} = \left( \sum_{i,j=1}^{3} A_{ij} N_i c_i^+ c_j \right) \underline{u} \tag{95}$$

By very similar calculations, which are far more tedious, we may verify that the $B_{ijk}$ terms have a similar effect, leading to the general result:
$$G_u = \sum_{i,j=1}^{n} A_{ij} N_i c_i^+ c_j + \sum_{i,j,k=1}^{n} B_{ijk} N_i c_i^+ c_j c_k \tag{96}$$

Note that the "A" term in equation 96 (or 94) creates an influence from terms of overall order $\ell+m+n$ to terms of the same order, while the "B" term creates an influence from terms of $\ell+m+n+1$ to terms of order

ℓ+m+n. This is the source of the infinite regress or closure of turbulence problem discussed above. All of the influence works from terms of some order to terms of equal or lower order. This is equivalent to saying that the annihilation operators outnumber the creation operators in some terms, but the creation operators do not outnumber the annihilation operators. In the picture of section 3.3.2, we would say that this gain matrix $G_u$ is lower triangular.

If we had introduced a quartic interaction into equation 89, based on a new term $\Sigma C_{ijkl}X_jX_kX_l$, essentially the same reasoning would have led to an additional term $\Sigma C_{ijkl}N_i c_i^+ c_j c_k c_l$ in equation 96. Clearly this would not have changed the qualitative conclusions here.

### 3.4.3 Dynamics of v for the ODE of Equations 3 and 89

From equation 10 and the definition of $N_i$ in section 3.2, we may write:

$$\underline{u} = (N_1!)^{1/2}(N_2!)^{1/2}(N_3!)^{1/2}\underline{v} \tag{97}$$

Substituting this into the equation $\partial_t \underline{u} = G_u \underline{u}$, with $G_u$ as in equation 96, and multiplying both sides by $N_1!^{-1/2} N_2!^{-1/2} N_3!^{-1/2}$ on the left, we may deduce:

$$\underline{\dot{v}} = \left( \sum_{i,j=1}^{n} A_{ij}(N_i^{1/2} c_i^+)(c_j N_j^{1/2}) + \sum_{i,j,k=1}^{n} B_{ijk}(N_i^{1/2} c_i^+)(c_j N_j^{1/2})(c_k N_k^{1/2}) \right) \underline{v} \tag{98}$$

The definitions in equations 34 and 35 tell us that we may write this more simply as:

$$\underline{\dot{v}} = G_v \underline{v} = \left( \sum_{i,j=1}^{n} A_{ij} a_i^+ a_j + \sum_{i,j,k=1}^{n} B_{ijk} a_i^+ a_j a_k \right) \underline{v} \tag{99}$$

This equation is somewhat simpler in appearance than equation 96, and slightly closer to the dynamics assumed in QFT. However, $G_v$ is still lower triangular, and the problem of infinite regress is still present.

### 3.4.4 Dynamics of w for the ODE of Equations 3 and 89

Our immediate goal here is to derive a dynamical equation for $\underline{w}$ which is valid for all physically realizable mixed states $\underline{w}$. We will do this in two different ways – a direct way, and an indirect way which illustrates the machinery of reification. Then we discuss some implications of the results.

In the direct approach, we first derive a dynamical equation for pure states $\underline{w}(\underline{X})$, again for the example of $\underline{X} \in R^3$. From equation 23, we may write:

$$\underline{w}(\underline{X}) = \exp(-\tfrac{1}{2}(X_1^2 + X_2^2 + X_3^2)) \underline{v}(\underline{X}) \tag{100}$$

Differentiating with respect to time, and then substituting in from equations 3 and 89, we get:

$$\begin{aligned}
\underline{\dot{w}}(\underline{X}) &= \exp\left(-\tfrac{1}{2}(X_1^2 + X_2^2 + X_3^2)\right)\underline{\dot{v}}(\underline{X}) - \left(X_1 \dot{X}_1 + X_2 \dot{X}_2 + X_3 \dot{X}_3\right) \exp\left(-\tfrac{1}{2}(X_1^2 + X_2^2 + X_3^2)\right)\underline{v} \\
&= \exp\left(-\tfrac{1}{2}(X_1^2 + X_2^2 + X_3^2)\right) G_v \underline{v}(\underline{X}) - \left(X_1 \dot{X}_1 + X_2 \dot{X}_2 + X_3 \dot{X}_3\right)\underline{w}(\underline{X}) \\
&= G_v \underline{w}(\underline{X}) - \left(\sum A_{ij} X_i X_j + \sum B_{ijk} X_i X_j X_k\right)\underline{w}(\underline{X})
\end{aligned}$$

Exploiting equation 51, we can simplify this to:

$$\underline{\dot{w}}(\underline{X}) = \left(G_v - \left(\sum A_{ij} a_i a_j + \sum B_{ijk} a_i a_j a_k \right)\right)\underline{w}(\underline{X})$$ (101)

Substituting in for $G_v$ from equation 99, we may write: (102)

$$G_w = \sum_{i,j=1}^{n} A_{ij}(a_i^+ - a_i)a_j + \sum_{i,j,k=1}^{n} B_{ijk}(a_i^+ - a_i)a_j a_k$$ (103)

Once again, the gain matrix is lower triangular, and we still have the problem of infinite regress or closure of turbulence. (By linearity, this $G_w$ applies to all physically realizable states.)

Our indirect derivation of $G_w$ is far more elaborate. We include it so as to give an illustration of the machinery of *reification*, which is crucial to the bridge to quantum theory.

The definition of $\underline{w}$ given in equations 78 and 75 may be written as:

$$\underline{w} = L\,\underline{y},$$ (104)

where:

$$L = S_{-}(-\tfrac{1}{2}) = \exp(-\tfrac{1}{2}\sum a_i a_i)$$ (105)

(Note that we have omitted the "$b_i$" operators for first-order ODE, as explained in section 3.3.6.)
In general, whenever we define a new vector $\underline{w}$ by an invertible linear transformation L of another vector $\underline{y}$ which obeys linear dynamics, we may deduce that:

$$\underline{\dot{w}} = L\underline{\dot{y}} = L G_v \underline{y} = L G_v L^{-1} \underline{w},$$

and thus: (106)

$$G_w = L G_v L^{-1}$$ (107)

The mathematician will immediately recognize this as a similarity transformation. (It is trivial to prove that the action of a similarity transformation on any polynomial expression can be derived by applying that similarity transformation directly to each of the matrices from which the polynomial is composed.) When $G_v$ is written as the sum of products of more elementary operators, we can calculate the results of such a similarity transformation simply by working out what it does to those elementary operators, and substituting them into $G_v$. In this case, the elementary operators are just the annihilation and creation operators, $a_k$ and $a_k^+$. Thus to derive $G_w = L G_v L^{-1}$, our first task is to work out $L a_k L^{-1}$ and $L a_k^+ L^{-1}$ for L as given in equation 105.

Because $a_k$ commutes with all the annihilation operators $a_i$ here, it also commutes with L as defined in equation 105. Thus:

$$L a_k L^{-1} = a_k L L^{-1} = a_k$$ (108)

In other words, the similarity transformation here does not change $a_k$ at all.

On the other hand, $a_k^+$ commutes with all of the annihilation operators $a_i$ *except* for $a_k$. Thus the terms $\exp(-\tfrac{1}{2}a_i a_i)$ for $i \neq k$ can be shifted to the right of $a_k^+$, when we work out $L a_k^+ L^{-1}$, and they cancel out their inverse terms on the right, just as happened for L in equation 108. Thus we are left with:

$$L a_k^+ L^{-1} = \exp(-\tfrac{1}{2}a_k a_k) a_k^+ (\exp(-\tfrac{1}{2}a_k a_k))^{-1}$$ (109)

Using the standard definition of a commutator (see section 3.3.2), we may write this as:

$$L a_k^+ L^{-1} = (\,[\exp(-\tfrac{1}{2}a_k a_k), a_k^+] + a_k^+ \exp(-\tfrac{1}{2}a_k a_k)\,)(\exp(-\tfrac{1}{2}a_k a_k))^{-1}$$

$$= [\exp(-\tfrac{1}{2}a_k a_k), a_k^+](\exp(-\tfrac{1}{2}a_k a_k))^{-1} + a_k^+$$ (110)

Using equation 42b, we may work out the commutator term in equation 110 as follows:

$$\left[\exp(-\tfrac{1}{2}a_k a_k), a_k^+\right] = \left[\sum_{i=0}^{\infty} \frac{1}{i!}(-\tfrac{1}{2})^i a_k^{2i}, a_k^+\right] = \sum_{i=0}^{\infty} \frac{1}{i!}(-\tfrac{1}{2})^i \left[a_k^{2i}, a_k^+\right] = \sum_{i=0}^{\infty} \frac{1}{i!}(-\tfrac{1}{2})^i (2i) a_k^{2i-1}$$

$$= a_k \sum_{i=0}^{\infty} -\frac{1}{(i-1)!}(-\tfrac{1}{2})^{i-1} a_k^{2i-2} = -a_k \exp(-\tfrac{1}{2}a_k a_k)$$

(111)

Substituting this back into equation 110, we discover that this complicating-looking similarity transformation merely transforms $a_k^+$ into $a_k^+ - a_k$!

Now that we know how the similarity transformation operates on $a_k$ and $a_k^+$, we can work out $G_w$ simply by going back to the expression for $G_v$ in equation 99, and replacing every annihilation operator $a_k$ by itself, and replacing every creation operator $a_k^+$ by $a_k^+ - a_k$. This immediately gives us the same expression for $G_w$ that appears in equation 103. This derivation is actually much easier than the direct derivation, *after* we understand how the similarity transformation works.

After we first derived $G_w$, we were slightly surprised and disappointed. $G_w$, like $G_u$ and $G_v$ is a lower triangular matrix. It is not antiHermitian. Yet **w** obeys linear dynamics, and its norm is always unchanged by these dynamics for all physically realizable states. (The norm is always 1 for those states.) There is a well-known theorem widely used in physics, as follows: if a vector **y** obeys $\partial_t \underline{y} = G_y \underline{y}$, and if the dynamics guarantee that $|\underline{y}|^2$ will always be unchanged over time, then $G_y$ must be antiHermitian. What happened here?

The explanation for this paradox is simple and inescapable. *IF* we could guarantee that the norm of **w** is unchanged by the dynamics, for all vectors **w** in Fock space, then $G_w$ would have to be antiHermitian. But we can guarantee this only for physically realizable states. Therefore, $G_w$ is allowed to be antiHermitian, only because this representation introduces the formal appearance of additional vectors **w** which are not physically realizable and whose norms are sometimes allowed to change over time.

The obvious next step here is to try to project **w** into the subspace of $F(R^n)$ spanned by pure states **w**(**X**). The projection of $G_w$ to that subspace would have to be antiHermitian. We have attempted this kind of approach, but we do not see a way to make it work in a useful way. The new results in section 3.3.7 suggest that we may need to do a projection from the matrix space, $F(R^n) \otimes F(R^n)$, to itself, rather than a simple projection from $F(R^n)$ to $F(R^n)$, in order to eliminate states which are not physically realizable.

### 3.4.5 Dynamics of Primal Matrix Representations ($\mathbf{r_u}, \mathbf{r_y}, \mathbf{r_w}$) For Eqs. (3) and (89)

Equations 53a defines $\rho_u(\underline{X})$ as $\underline{u}(\underline{X})\underline{u}^H(\underline{X})$. From this we easily deduce:

$$\partial_t \rho_u(\underline{X}) = (\partial_t \underline{u}(\underline{X}))\underline{u}^H(\underline{X}) + \underline{u}(\underline{X})(\partial_t \underline{u}(\underline{X}))^H = G_u \underline{u}(\underline{X})\underline{u}^H(\underline{X}) + \underline{u}(\underline{X})\underline{u}^H(\underline{X}) G_u^H$$

$$= G_u \rho_u(\underline{X}) + \rho_u(\underline{X}) G_u^H \qquad (112)$$

Taking expectation values on both sides, we deduce, for all physically realizable states:

$$\partial_t \rho_u = G_u \rho_u + \rho_u G_u^H \qquad (113)$$

Thus the gain matrix for $\rho_u$, as defined in equation 91, is just $G_u$, the same gain matrix as for **u** itself! Using identical logic, starting from the definitions in equations 53b and 53c, we easily deduce that the gain matrix for $\rho_v$ is just $G_v$, and that the gain matrix for $\rho_w$ is $G_w$.

### 3.4.6. Dynamics of $\underline{b}_u$ for ODE Given in Eqs (3) and (89)

This section will derive the gain matrix $G_{bu}$ which governs the vector $\underline{b}_u$ defined in equation 69.

As with **w**, we will give two derivations – one direct and one indirect – which provide a check on each other, and also provide some theoretical insight. But in this case, the direct derivation will be far more tedious than the indirect derivation.

The direct derivation begins by considering the example where n=3, where the definition of **b**$_u$ in equation 69 is simply:

$$p(\underline{X}) = \sum_{i,j,k=0}^{\infty} b_{ijk} X_1^i X_2^j X_3^k \qquad (114)$$

In order to avoid confusion with indices, we first rewrite the Fokker-Planck equation (5) and our choice for **f** (equation 89) with different indices:

$$\dot{p} + \sum_{l=1}^{3} \frac{\partial}{\partial l} \left( p(\underline{X}) f_l(\underline{X}) \right) = 0 \qquad (115)$$

$$f_l(\underline{X}) = \sum_{m=1}^{3} A_{lm} X_m + \sum_{m,n=1}^{3} B_{lmn} X_m X_n \qquad (116)$$

As with our analysis of the dynamics of **u**, it is easiest to begin by considering the special case where $B_{\lambda\mu\nu}=0$, where:

$$f_\lambda = A_{\lambda 1} X_1 + A_{\lambda 2} X_2 + A_{\lambda 3} X_3 \qquad (117)$$

Multiplying equation 114 by 117, we get:

$$pf_1 = \sum_{i,j,k=0}^{\infty} b_{ijk} \left( A_{11} X_1^{i+1} X_2^j X_3^k + A_{12} X_1^i X_2^{j+1} X_3^k + A_{13} X_1^i X_2^j X_3^{k+1} \right) \qquad (118)$$

If we consider the case where $\lambda=1$, for example, we may differentiate to get:

$$\frac{\partial}{\partial X_1}(pf_1) = \sum_{i,j,k=0}^{\infty} b_{ijk} \left( A_{11}(i+1) X_1^i X_2^j X_3^k + A_{12}(i) X_1^{i-1} X_2^{j+1} X_3^k + A_{13}(i) X_1^{i-1} X_2^j X_3^{k+1} \right) \qquad (119)$$

If we substitute this expression, plus the equivalent expression for $\lambda=2$ and $\lambda=3$, plus our definition in equation 114, into equation 115, and take the expectation value, we obtain:

$$0 = \sum_{i,j,k=0}^{\infty} \left\{ \dot{b}_{ijk} u_{ijk} + b_{ijk} \begin{pmatrix} (i+1)A_{11} u_{ijk} + iA_{12} u_{i-1,j+1,k} + iA_{13} u_{i-1,j,k+1} \\ + jA_{21} u_{i+1,j-1,k} + (j+1)A_{22} u_{ijk} + jA_{23} u_{i,j-1,k+1} \\ + kA_{31} u_{i+1,j,k-1} + kA_{32} u_{i,j+1,k-1} + (k+1)A_{33} u_{ijk} \end{pmatrix} \right\} \qquad (120)$$

Recall that we want to find values for $\{\partial_t b_{ijk}\}$ such that this sum will always be zero, for all possible values of $\{u_{ijk}\}$. (Actually, it would be enough to make it zero for all physically realizable values of **u**, but it is not easy to exploit the extra freedom which this offers.) In order to do this, we must re-express this sum by collecting all terms which multiply each element $u_{ijk}$. Formally, we do this by considering equation 120 to be the sum of ten separate sums, and redefining the indices in each sum so that the "u" factor appears as $u_{ijk}$. This results in:

$$0 = \sum_{i,j,k=0}^{\infty} u_{ijk} \left\{ \begin{array}{l} \dot{b}_{ijk} + b_{ijk}\left((i+1)A_{11} + (j+1)A_{22} + (k+1)A_{33}\right) \\ + b_{i+1,j-1,k}(i+1)A_{12} + b_{i+1,j,k-1}(i+1)A_{13} \\ + b_{i-1,j+1,k}(j+1)A_{21} + b_{i,j+1,k-1}(j+1)A_{23} \\ + b_{i-1,j,k+1}(k+1)A_{31} + b_{i,j-1,k+1}(k+1)A_{32} \end{array} \right\}$$

(121)

In interpreting this sum, it is important to realize that elements $b_{ijk}$ for which i, j or k are less than zero should be interpreted as zero. In order to make sure that this sum is zero, for all possible sets of values $\{u_{ijk}\}$, we simply require that:

$$-\dot{b}_{ijk} = b_{ijk}\left((i+1)A_{11} + (j+1)A_{22} + (k+1)A_{33}\right) + b_{i+1,j-1,k}(i+1)A_{12} + b_{i+1,j,k-1}(i+1)A_{13}$$
$$+ b_{i-1,j+1,k}(j+1)A_{21} + b_{i,j+1,k-1}(j+1)A_{23} + b_{i-1,j,k+1}(k+1)A_{31} + b_{i,j-1,k+1}(k+1)A_{32}$$

(122)

again with the understanding that elements of $b_{ijk}$ with i or j or k less than zero are treated as zero.

Equation 122 already gives us the required dynamics of $\underline{b}_u$, in principle. But we wish to express the dynamics in a more elegant form, exploiting our definitions for operators over Fock space. This is somewhat tricky. For example, consider the term "$b_{i+1,j-1,k}(i+1)A_{12}$". This term "annihilates" an $X_1$ factor, sending i+1 to i. It "creates" an $X_2$ factor. The overall effect on $\underline{b}_u$ is $A_{12}c_1c_2^+N_1$, which equals $A_{12}(N_1+1)c_1c_2^+$. Similar analysis of the various terms in equation 122 show that equation 122 can be expressed as:

$$\dot{\underline{b}}_u = -\left(\sum_{i,j=1}^{3} A_{ij}(N_i+1)c_i c_j^+\right) \underline{b}_u$$

(123)

We have performed this same set of calculations for the general case, where B may not equal zero. They are extremely tedious. They simply result in confirming the obvious extrapolation:

$$\dot{\underline{b}}_u = -\left(\sum_{ij} A_{ij}(N_i+1)c_i c_j^+ + \sum_{ijk} B_{ijk}(N_i+1)c_i c_j^+ c_k^+\right) \underline{b}_u$$

(124)

Note that the gain matrix here is completely upper triangular, in the picture of section 3.3.2. The number of creation operators in every term is sometimes greater than the number of annihilation operators and sometimes equal. This is exactly the opposite of the situation with $\underline{u}$ itself! In fact, the gain matrix here is almost the Hermitian conjugate or adjoint of $G_u$, multiplied by (-1). This is what one might expect from the general appearance of duality between $\underline{u}$ and $\underline{b}_u$ in equation 69.

Our second derivation of $G_{bu}$ is based on a direct exploitation of this duality. However, one must be careful here. The relationship between $\underline{b}_u$ and $\underline{u}$ is very subtle. (Also, in performing such calculations, the novice should be reminded, for example, that $c_i c_j^+$ is not equivalent to $c_j^+ c_i$.)

In order to understand this relationship, it helps to go back to the original form of the Fokker-Planck equation, as it is used to describe the dynamics of compressible fluids. In that application, $p(\underline{X})$ is actually written as $\rho(\underline{x})$, and it represents the density of fluid molecules at some point in three-dimensional space. Thus the Fokker-Planck equation is sometimes written as:

$$\partial_t \rho + (\nabla \rho \cdot \underline{v}) + \rho (\text{div } \underline{v}) = 0 \qquad (125)$$

The first term in the equation represents the change of density at some point $\underline{x}$ in space. Intuitively, the density of fluid at any point $\underline{x}$ changes as the result of two factors: (1) the current of flow, $\underline{v}$, is sending fluid from some other point, $\underline{x}_0$, to the point $\underline{x}$; this changes the density at $\underline{x}$ towards equaling the density at $\underline{x}_0$; (2) various factors, like walls squeezing the flow, may actually compress the flow, even along a given streamline. More formally, it is possible to analyze the change in density at point **x** into the sum of two terms, one which equals the change in density along a given streamline, and one which accounts for motion along the streamline. The last term in equation 125 represents the compression effect, the change in density along a streamline. The middle term represents the effect of motion along a streamline.

Now consider the time-varying function P(t) defined by:

$$P(t) = \underline{b}_u(t)^H \underline{u}(t) \tag{126}$$

This function is very similar to equation 69, but be careful! If we start out in a statistical state $\underline{u}(0) = \underline{u}(\underline{X}(0))$ based on a physical state $\underline{X}(0)$, then $\underline{u}(t)=\underline{u}(\underline{X}(t))$ will incorporate the changes in $\underline{X}$ over time. Thus in equation 126 we are not measuring the change in probability at a fixed point $\underline{X}$. We are tracking the change in probability density along a streamline in the space $R^n$. By differentiating equation 126 with respect to time, we get:

$$\partial_t P(t) = \partial_t (\underline{b}_u(t)^H \underline{u}(t)) = \underline{b}_u(t)^H G_{bu}^H \underline{u}(t) + \underline{b}_u(t)^H G_u \underline{u}(t)$$

$$= \underline{b}_u^H (G_{bu}^H + G_u) \underline{u} \tag{127}$$

But because $\partial_t P(t)$ is the change in density along a streamline, equation 125 with the streamline removed (and returning to our notation from equation 3) gives us:

$$\partial_t P(t) = -p \sum_i \frac{\partial f_i}{\partial X_i} \tag{128}$$

From equation 89, we may easily calculate:

$$\sum_i \frac{\partial f_i}{\partial X_i} = \text{Tr } A + \sum_{i,k} (B_{iik} + B_{iki}) X_k \tag{129}$$

For any pure state $\underline{u}(\underline{X})$, we may combine equations 69, 128 and 129 to deduce:

$$\partial_t P(t) = -\underline{b}_u^H \underline{u}(\underline{X}) \left( \text{Tr } A + \sum_{i,k} (B_{iik} + B_{iki}) X_k \right) = -\underline{b}_u^H \left( \text{Tr } A + \sum_{i,k} (B_{iik} + B_{iki}) c_k \right) \underline{u}(\underline{X}) \tag{130}$$

Taking the expectation value of this, and combining it with equation 127, we deduce:

$$G_{bu} = -\left( G_u + \text{Tr } A + \sum_{i,k} (B_{iik} + B_{iki}) c_k \right)^H = -G_u^H - \text{Tr } A - \sum_{i,k} (B_{iik} + B_{iki}) c_k^+ \tag{131}$$

In effect, $-G_u^H$ gives us the stream term in the Fokker-Planck equation, while the other two terms give us the compression term. Substituting in from equation 96, this gives a slightly more convenient (albeit equivalent) description of the dynamics of $\underline{b}_u$:

$$\underline{\dot{b}}_u = \left( -\sum_{i,j} A_{ij} c_j^+ (c_i N_i) - \sum_{i,j,k} B_{ijk} c_j^+ c_k^+ (c_i N_i) - \text{Tr } A - \sum_{i,k} (B_{iik} + B_{iki}) c_k^+ \right) \underline{b}_u \tag{132}$$

### 3.4.7. Dynamics of $\underline{b}_v$ for ODE Given in Eqs (3) and (89)

We can derive the dynamics of $\underline{b}_v$ simply by expressing $\underline{b}_u$ as a function of $\underline{b}_v$, and substituting back into the dynamics of $\underline{b}_u$ as given in equation 132.

First consider the definition of $\underline{b}_v$ for the case n=3:

$$p = \underline{b}_v^H \underline{v} = \sum_{i,j,k} b_{ijk}^{[v]} v_{ijk} = \sum_{i,j,k} b_{ijk}^{[v]} \frac{X_1^i X_2^j X_3^k}{\sqrt{i!j!k!}} \tag{133}$$

which must equal:

$$\sum_{i,j,k} b_{ijk}^{[u]} X_1^i X_2^j X_3^k \tag{134}$$

Comparing these two equations, we deduce:

$$\underline{b}_u = (N_1!)^{-1/2}(N_2!)^{-1/2}(N_3!)^{-1/2} \underline{b}_v \tag{135}$$

Substituting equation 135 into equation 132, and multiplying both sides on the left by $(N_1!)^{1/2}(N_2!)^{1/2}(N_3!)^{1/2}$, we may derive:

$$\dot{\underline{b}}_v = \left(-\sum_{i,j} A_{ij}(N_j^{1/2} c_j^+)(c_i N_i^{1/2}) - \sum_{i,j,k} B_{ijk}(N_j^{1/2} c_j^+)(N_k^{1/2} c_k^+)(c_i N_i^{1/2}) - \text{Tr } A - \sum_{i,k}(B_{iik} + B_{iki})(N_k^{1/2} c_k^+)\right) \underline{b}_v$$

$$= -\left(\sum_{i,j} A_{ij} a_j^+ a_i + \sum_{i,j,k} B_{ijk} a_j^+ a_k^+ a_i + \text{Tr } A + \sum_{i,k}(B_{iik} + B_{iki}) a_k^+\right) \underline{b}_v \tag{136}$$

As with $\underline{b}_u$, the gain matrix is totally upper triangular. There is no infinite regress problem at all.

### 3.4.8. Dynamics of $\underline{b}_w$ for ODE Given in Eqs (3) and (89)

To derive the dynamics of $\underline{b}_w$ in this system, we can simply apply duality exactly as we did with $\underline{b}_u$ in section 3.4.6. In this case, the quantities $X_i$ from the compression term will yield the operators $a_i$ instead of $c_i$, because of equation 51. Other than that, the derivation is the same. Thus in parallel to equation 131, we obtain:

$$G_{bw} = -\left(G_w + \text{Tr } A + \sum_{i,k}(B_{iik} + B_{iki}) a_k\right)^H$$

$$= -\sum_{i,j} a_j^+(a_i - a_i^+) - \sum_{i,j,k} B_{ijk} a_k^+ a_j^+(a_i - a_i^+) - \text{Tr } A - \sum_{i,k}(B_{iik} + B_{iki}) a_k^+ \tag{137}$$

As with $\underline{b}_u$ and $\underline{b}_v$, the gain matrix is again upper triangular.

### 3.4.9. Dynamics of Entropy Vectors $\underline{b}_u$, $\underline{b}_v$, $\underline{b}_w$ for ODE of Eqs (3) and (89)

These three dual representations of Pr(state) appear to have no intrinsic computational value. However, because they are useful as an intermediate step in our discussion, we will redefine them here, and derive their dynamics.

The entropy vector $\underline{b}_u$ represents Pr($\underline{X}$) as:

$$S(\underline{X}) = \log \Pr(\underline{X}) = \underline{b}_u{}^H \underline{u}(\underline{X}) \tag{138}$$

$\underline{b}_v$ and $\underline{b}_w$ are defined in the obvious analogous way.

To derive the dynamics of $\underline{b}_u$, we must first develop a dynamical equation for the entropy S. To do this, we first write

$$p = e^S, \tag{139}$$

and substitute into the Fokker-Planck equation, which yields:

$$e^S \dot{S} + \sum_i e^S \frac{\partial S}{\partial X_i} f_i + e^S \sum_i \frac{\partial f_i}{\partial X_i} = 0, \tag{140}$$

which reduces to:

$$\dot{S} + \sum_i \frac{\partial S}{\partial X_i} f_i + \sum_i \frac{\partial f_i}{\partial X_i} = 0 \tag{141}$$

Comparing this with the Fokker Planck equation, we see something truly remarkable: the equation is exactly the same, with the same stream term, *except that* the compression term is now a kind of exogenous "source term" which does not depend at all on S itself! If we go back to equation 131, and see what gain matrix we would have had without a compression term, and then go back and see how the direct derivation of $G_{bu}$ would have been influenced by this exogenous source term, we immediately derive:

$$\partial_t \underline{b}_u = -G_u{}^H \underline{b}_u - \underline{s}^{[u]}, \tag{142}$$

where $\underline{s}^{[u]}$ is that vector $\underline{s}$ in Fock space for which:

- $s_0 = \text{Tr } A$
- $(s_1)_k = \Sigma (B_{iik} + B_{iki})$
- $s_k = 0$ for all $k>1$

Likewise we may deduce that:

$$\partial_t \underline{b}_v = -G_v{}^H \underline{b}_v - \underline{s}^{[v]}, \tag{143}$$

$$\partial_t \underline{b}_w = -G_w{}^H \underline{b}_w - \underline{s}^{[w]}, \tag{144}$$

All of these entropy vectors have upper triangular gain matrices, and all three of them have exogenous "source" terms $\underline{s}$. These exogenous terms are radically different from what we see in quantum theory, but they are an important part of generalized statistics or thermodynamics.

There are many dynamical systems with coefficients A and B which would happen to have source terms $\underline{s}$ identically equal to zero. (For example, the second-order ODE which we will consider here have an A matrix with a term $-im_k$ to balance out every $+im_k$ term, leading to a trace of zero.) We would call such systems "noncompressive" or "statistically noncompressive" systems. We would tend to expect that the field theories addressed in current QFT are statistically noncompressive. For statistically noncompressive dynamical systems, the dynamical equations for $\underline{b}_u, \underline{b}_v$ and $\underline{b}_w$ are identical to those for $\underline{b}_u, \underline{b}_v$ and $\underline{b}_y$, with gain matrices exactly adjoint (with a minus sign) to the gain matrices for $\underline{u}, \underline{v}$ and $\underline{w}$.

**3.4.10. Dynamics of $B_w$ for ODE Given in Eqs (3) and (89)**

In section 3.3.4, we briefly mentioned the representations $B_u$, $B_v$, $B_w$ and $B_z$. These are the dual matrix representations, parallel to the dual vector representations $\underline{b}_u$, $\underline{b}_v$, $\underline{b}_w$ and $\underline{b}_z$. These representations do not seem so useful for their own sake. However, they do provide an excellent starting point for the analysis of entropy matrices discussed in section 3.4.11.

This subsection will derive the dynamics of $B_w$, based on the same duality approach used in the later part of section 3.4.6 in order to derive the dynamics of $\underline{b}_u$. Here we will merely sketch out what is different for $B_w$.

The vector representation $\underline{b}_u$ was defined in equation 69. For $B_w$, instead of equation 69, we have the definition:

$$\Pr(\underline{X}) = \tfrac{1}{2}\,\underline{w}(\underline{X})^H B_w \underline{w}(\underline{X}) \tag{145}$$

In parallel with equation 126, we may define the probability density P along a streamline:

$$P(t) = \tfrac{1}{2}\,\underline{w}(t)^H B_w(t)\underline{w}(t) \tag{146}$$

We may again differentiate with respect to time, to get (for pure states $\underline{w}$):

$$\partial_t P(t) = \tfrac{1}{2}\,\underline{w}(t)^H G_w^H B_w(t)\underline{w}(t) + \tfrac{1}{2}\,\underline{w}(t)^H B_w(t) G_w \underline{w}(t) + \tfrac{1}{2}\,\underline{w}(t)^H (\partial_t B_w)\underline{w}(t)$$

$$= \tfrac{1}{2}\,\underline{w}^H\,(G_w^H B_w + B_w G_w + \partial_t B_w)\,\underline{w} \tag{147}$$

In place of equation 130, we may use equation 51 and its Hermitian conjugate to deduce:

$$\partial_t P(t) = -p\sum_i \frac{\partial f_i}{\partial X_i} = -\underline{w}(\underline{X})^H B_w \underline{w}(\underline{X})\left(\operatorname{Tr} A + \sum_{i,k}(B_{iik}+B_{iki})X_k\right)$$

$$= -\tfrac{1}{2}\underline{w}(\underline{X})^H\left(\left(\operatorname{Tr} A + \sum_{i,k}(B_{iik}+B_{iki})a_k^+\right) B_w + B_w\left(\operatorname{Tr} A + \sum_{i,k}(B_{iik}+B_{iki})a_k\right)\right) \tag{148}$$

Combining equations 147 and 148, we satisfy the requirements across all possible vectors $\underline{w}(\underline{X})$ by setting:

$$\partial_t B_w = -\left(G_w^H + \operatorname{Tr} A + \sum_{i,k}(B_{iik}+B_{iki})a_k^+\right) B_w - B_w\left(G_w + \operatorname{Tr} A + \sum_{i,k}(B_{iik}+B_{iki})a_k\right) \tag{149}$$

Notice that this derivation "leaves the mixing to $B_w$"; in other words, it is sufficient that equation 149 meets the requirement of equation 148 for all pure states $\underline{w}(\underline{X})$. Equation 145 provides the machinery for using $B_w$ itself to represent a probability distribution for a mix of such pure states.

Using the definition of "gain matrix" in equation 91, we have just shown that the gain matrix for $B_w$ is in fact identical to the gain matrix for $\underline{b}_w$, $G_{bw}$, given in equation 137. We would expect that $G_{Bu}=G_{bu}$, $G_{Bv}=G_{bv}$ and $G_{Bz}=G_{bz}$ as well, based on exactly the same logical parallels. (In parallel with earlier discussions, note that equation 149 does not give the one and only unique version of $\partial_t B_w$ which leads to the correct probability distribution for *physically realizable states*. See the discussion of section 3.3.4 concerning the nonuniqueness of dual representations. It is sufficient for our purposes that equation 149 gives *one* of the paths for $B_w$ which represents the correct probability distribution.)

### 3.4.11. Dynamics of the Entropy Matrix $S_w$ For Eqs (3) and (89)

Section 3.3.4 argued that the entropy matrix $S_w$ is our best hope for "closing turbulence" and performing statistical analysis of a wide range of dynamical systems. (By contrast, reification is currently our best mechanism for making contact with existing quantum theory.)

We have already done most of the work required to derive the dynamics of $S_w$. In section 3.4.9, we observed that the dynamic equation for entropy, equation 141, is almost identical to the Fokker-Planck equation, equation 5. Thus we found that the dynamic equation for $\mathbf{b}_u$ is based on the same gain matrix as $\mathbf{b}_u$, *but without* the compression term; that in turn is the same as $-G_u^H$, where $G_u$ is the gain matrix for $\mathbf{u}$.

The exact same logic applies to $S_w$. Thus in parallel with section 3.4.9, we can reasonably extrapolate equations 144 and 149 to deduce:

$$\partial_t S_w = -G_w^H S_w - S_w G_w - S^{[w]}, \qquad (150)$$

where $S^{[w]}$ is an exogenous source matrix, a kind of matrix constant term. Likewise, we would expect:

$$\partial_t S_u = -G_u^H S_u - S_u G_u - S^{[u]}, \qquad (151)$$

$$\partial_t S_v = -G_v^H S_v - S_v G_v - S^{[v]}, \qquad (152)$$

$$\partial_t S_z = -G_z^H S_z - S_z G_z - S^{[z]}, \qquad (153)$$

The dynamics of these matrices are all upper triangular, in the picture of section 3.3.2.
Thus there is no infinite regress or closure of turbulence problem.
As in our previous discussions of duality, we would expect that the exogenous source term would be zero for all the PDE immediately relevant to quantum physics, which appear to be statistically noncompressive. Thus we have not had reason to perform the exact, careful calculation needed to work it out exactly. However, it is easy to see the outlines of one way to do this calculation, at least for $S^{[u]}$.

The calculation would begin by writing out a *direct* derivation of the dynamics of $B_u$, just like the direct derivation of $\mathbf{b}_u$ in section 3.4.6. It would then require that a very complex sum be set to zero, as in equation 120. But we would add a term representing the expected value of $(\text{Tr}A + \Sigma(B_{iik}+B_{iki})X_k)$; this would add a few terms to that sum of the form $(\partial_t S_{set1;set2})u_{set1}u_{set2}$. These terms would be nonzero only for cases where one of the two sets of indices is "000" and the other is "000," "100," "010," or "001." The net effect would be to gives us an exogenous source matrix $S^{[u]}$ of the form

$$k\,|0\rangle\langle 0| + |0\rangle\underline{\mathbf{s}}^H + \underline{\mathbf{s}}\langle 0| \qquad (154)$$

where "$\langle 0|$" is the transpose of $|0\rangle$, where k is a scalar, and $\underline{\mathbf{s}}$ is a low-order source vector very similar to the source vectors of section 3.4.9. Again, however, the remainder of this paper will focus mainly on statistically noncompressive systems, for which the exogenous source term is zero.

Because there are many different matrices $S_w(t)$ which can represent the same probability distribution $\text{Pr}(\underline{\mathbf{X}})$, it seems possible that equation 150 could be valid *and that* an alternative version of equation 150 could *also* be valid, using some other matrix A instead of $G_w$. In other words, two different dynamical equations giving different predictions of $\partial_t S_w$ could both be valid, because they could point to different matrices $S_w(t+dt)$ which correspond to the *same* probability distribution $\text{Pr}(\underline{\mathbf{X}}(t+dt))$. This would essentially require that the matrix $\delta = G_w - A$ have the property that $\delta S_w$ is a null operator, as defined in section 3.3.7, for *all* admissible matrices $S_w$. So far as we can tell, this is impossible, but the analysis is rather complicated.

### 3.4.12. Reification of the First-Order ODE of Eqs (3) and (89)

This section will discuss the dynamics of the vector $\underline{z}$ defined in section 3.6 for this case:

$$\underline{z} = S_-(-\tfrac{1}{2})S_+(-\tfrac{1}{4})\underline{v} = \exp\left(-\frac{1}{2}\sum_{i=1}^{n} a_i a_i\right)\exp\left(-\tfrac{1}{4}\sum_{i=1}^{n} a_i^+ a_i^+\right)\underline{v}$$

(155)

We will find that the gain matrix for $\underline{z}$, $G_z$, is antiHermitian, for systems which are statistically noncompressive (as defined in section 3.4.6). Therefore, by equation 153, the gain matrix for the reified entropy matrix, $S_z$, is also equal to $G_z$ for those systems.

In order to derive $G_z$, we will use the same approach given in the later part of section 3.4.4. In parallel with equation 107, we have:

$$G_z = L G_v L^{-1} = L_2 L_1 G_v L_1^{-1} L_2^{-1}, \quad (156)$$

where:

$$L = L_2 L_1 = \exp(-\tfrac{1}{2}\sum_{i=1}^{n} a_i a_i)\exp(-\tfrac{1}{4}\sum_{i=1}^{n} a_i^+ a_i^+)$$

(157)

This tells us that we can derive $G_z$ by performing *two* similarity transformations on $G_v$. First we perform $L_1$ and then we perform $L_2$. In other words, the overall similarity transformation L is defined by considering what happens if we first perform $L_1$ and then perform $L_2$.

In section 3.4.4, we have already derived what the similarity transform $L_2 = \exp(-\tfrac{1}{2}\Sigma a_i a_i)$ does. It transforms every annihilation operator $a_k$ into itself. It transforms every creation operator $a_k^+$ into $a_k^+ - a_k$. But what does $L_1$ do? $L_1$ transforms every creation operator into itself, because $L_1$ commutes with every creation operator. $L_2$ also transforms every annihilation operator $a_k$ into something analogous to equation 109:

$$a_k \;\to\; \exp(-\tfrac{1}{4}\, a_k^+ a_k^+)\, a_k\, (\exp(-\tfrac{1}{4}\, a_k^+ a_k^+))^{-1}$$

$$= (\,[\exp(-\tfrac{1}{4}\, a_k^+ a_k^+), a_k] + a_k \exp(-\tfrac{1}{4}\, a_k^+ a_k^+)\,)\, (\exp(-\tfrac{1}{4}\, a_k^+ a_k^+))^{-1}$$

$$= [\,\exp(-\tfrac{1}{4}\, a_k^+ a_k^+), a_k]\, (\exp(-\tfrac{1}{4}\, a_k^+ a_k^+))^{-1} + a_k \quad (158)$$

In parallel with equation 111, we can work out the commutator here by exploiting equation 42a:

$$\left[\exp(-\tfrac{1}{4} a_k^+ a_k^+), a_k\right] = \left[\sum_{i=0}^{\infty}\frac{1}{i!}(-\tfrac{1}{4})^i (a_k^+)^{2i}, a_k\right] = \sum_{i=0}^{\infty}\frac{1}{i!}(-\tfrac{1}{4})^i [(a_k^+)^{2i}, a_k] = \sum_{i=0}^{\infty}\frac{1}{i!}(-\tfrac{1}{4})^i(-2i)(a_k^+)^{2i-1}$$

$$= \tfrac{1}{2} a_k^+ \sum_{i=0}^{\infty}\frac{1}{(i-1)!}(a_k^+)^{2i-2}(-\tfrac{1}{4})^{i-1} = \tfrac{1}{2} a_k^+ \exp(-\tfrac{1}{4} a_k^+ a_k^+)$$

(159)

Substituting equation 159 into equation 158, we immediately see that $L_1$ simply transforms $a_k$ into $a_k + \tfrac{1}{2} a_k^+$.

Now let us see what the sequence of similarity transformations $L = L_2 L_1$ does to $a_k^+$ and $a_k$. First let us consider $a_k^+$. $L_1$ simply transforms $a_k^+$ into itself. Then $L_2$ transforms $a_k^+$ into $a_k^+ - a_k$. Thus the overall sequence, $L = L_2 L_1$, converts $a_k^+$ into $a_k^+ - a_k$. Next let us consider $a_k$. $L_1$ converts $a_k$ into $a_k + \tfrac{1}{2} a_k^+$. $L_2$ then converts this into $a_k + \tfrac{1}{2}(a_k^+ - a_k)$ which equals $\tfrac{1}{2}(a_k + a_k^+)$. In summary, the overall similarity transform L has the effect:

$$a_k^+ \;\to\; a_k^+ - a_k$$

$$a_k \;\to\; \tfrac{1}{2}(a_k + a_k^+)$$

(160)

In order to derive $G_z$, we simply apply this similarity transformation directly to $G_v$, as we did with the simpler similarity transformation in section 3.4.4. Simply by applying equation 160 to $G_v$ as given in equation 99, we derive:

$$G_z = \tfrac{1}{2}\sum_{i,j=1}^{n} A_{ij}(a_i^+ - a_i)(a_j^+ + a_j) + \tfrac{1}{4}\sum_{i,j,k=1}^{n} B_{ijk}(a_i^+ - a_i)(a_j^+ + a_j)(a_k^+ + a_k)$$

(161)

At first glance, this matrix appears to be antiHermitian. Certainly all terms of this sum are antiHermitian for the cases where i≠j (with A) and i≠j≠k≠i (with B). But an exact calculation shows that $[G_z, G_z^H]$ is proportional to the compression term given in equation 129. Thus $G_z$ is antiHermitian if and only if the dynamical system defined by equations 3 and 89 is statistically noncompressive.

In the noncompressive case, where $G_z$ is antiHermitian and $S^{[z]}=0$, the dynamic equations for **z** and for $S_z$ (equation 153) are essentially the same as the old familiar equations 1 and 2 of quantum theory. Those equations preserve the $L^2$ norms of **z** and $S_z$. Because of the way in which probabilities and expectation values are calculated in quantum situations, there is no loss of generality in simply normalizing these quantities to one, when those tools are being used. The resulting "normalized $S_z$", $S_z/Tr(S_z)$, appears identical to the usual density matrix of modern quantum theory in most respects.

We can imagine a possibility that $S_z$ or $S_z/Tr(S_z)$ might be defined in a different but equivalent way. Perhaps they could be defined as the result of "normalizing" the matrix $S_w$ to unit trace, or something similar. Recall that **w** itself was defined as the result of normalizing the vector **y**. For the canonical version of $S_w$, as defined in section 3.3.7, the matrix $S_w/Tr(S_w)$ may be represented as something like $M_-()S_w M_+()$, analogous to the alternate definition of **w** in equation 78. If the resulting normalized matrix obeys antiHermitian dynamics, then perhaps these dynamics would continue to represent the canonical version of $S_w/Tr(S_w)$. We have not yet explored these kinds of possibilities in detail, because it is enough for now to know that we have at least ONE set of definitions for $S_z$ and $S_z/Tr(S_z)$ which mirror quantum theory. However, a deeper understanding of reification and of alternative approaches to reification could be very valuable in the future.

## 3.5 Dynamics of Representations of Pr(X) for Second-Order ODE

### 3.5.1 Goals and Conclusions

The previous section provided excruciating detail in deriving the dynamics of about twenty different possible representations of the probability distribution Pr(**X**), for first-order ODE. Some of these representations, like the entropy matrix $S_w$, could also be very useful in the analysis of second-order ODE and PDE.

The goals of this section are more limited. Our work on second-order systems has been motivated by two major goals:

o To see how much the overall pattern in section 3.4 is changed when we go to the case
  of second-order systems or to systems where **X**∈$C^n$ and where there is some kind of
  symmetry between **X** and **X**-conjugate.

o To analyze the dynamics of **y** and **z** for the specific second-order ODE which correspond closely
  to the nonlinear Klein-Gordon equation (15), including the
  famous special cases where f(φ) is $φ^2$ or $φ^3$.

Section 3.5.2 will discuss the general case of systems:

$$\partial_t^2 \mathbf{X} = \mathbf{f}(\mathbf{X}),$$ (162)

where **X**∈$R^n$ and **f** is analytic. These systems can be represented by the equivalent first-order ODE whose state **Y**∈$R^{2n}$ is simply $\mathbf{X} \oplus (\partial_t \mathbf{X})$. These systems are always statistically noncompressive, as defined in section 3.4.9. This implies that the $S_w$ matrix obeys equation 150, with no exogenous source terms (i.e. with $S^{[w]}=0$). But we will not derive $G_w$, since it is just a special case of section 3.4. We *will* derive $G_y$ and $G_z$, because of the relation to the nonlinear Klein-Gordon equation. We will also discuss how energy conservation works in such systems. We will assume a familiarity with section 3.4, and will not repeat all the intermediate calculations.

Section 3.5.3 will discuss the special case of second-order ODE described by equation 11. It will show how the modified representations given in equation 12, etc., yield an alternative version of $G_y$ and $G_z$ which is closer to what is used in quantum field theory, in analyzing similar special cases.

The general results of sections 3.5.2 and 3.5.3 can be applied to the special case where $\underline{X} \in R^1$ and where $\underline{f}(\underline{X})$ is just $X^2$ or $X^3$. The two versions of $G_y$ and $G_z$ derived in these sections will both require the use of *four* creation and annihilation operators in this case $(a,b,a^+,b^+)$. This will also turn out to be true for the full nonlinear Klein-Gordon equation, to be discussed in section 3.6. But in quantum field theory, only one pair of operators $(a,a^+)$ is needed to describe real scalar Klein-Gordon theory. Section 4 will attempt to explain and overcome this discrepancy.

### 3.5.2. Statistical Dynamics and Energy Conservation For Systems Governed by Equation 162

Let us define the vector $\underline{Y} \in R^{2n}$ as $\underline{X} \oplus (\partial_t \underline{X})$. With that definition, equation 162 is equivalent to the first-order system:

$$\partial_t \underline{Y} = \partial_t \begin{bmatrix} Y_1,...,Y_n \\ Y_{n+1},...,Y_{2n} \end{bmatrix} = \partial_t \begin{bmatrix} \underline{X} \\ \partial_t \underline{X} \end{bmatrix} = \begin{bmatrix} \partial_t \underline{X} \\ \underline{f}(\underline{X}) \end{bmatrix} = \underline{F}(\underline{Y})$$

(163)

where we define:

$$\begin{aligned} F_i(\underline{Y}) &= Y_{i+n} \quad &&\text{(i.e., } (\partial_t \underline{X})_i) \\ &&& (1 \le i \le n) \\ F_{i+n}(\underline{Y}) &= f_i(Y_1,...,Y_n) \quad &&\text{(i.e., } f_i(\underline{X})) \end{aligned}$$

(164)

From section 3.4.9, we know that the first-order system given in equation 163 is statistically noncompressive if and only if:

$$\sum_{i=1}^{2n} \frac{\partial F_i}{\partial Y_i} = 0$$

(165)

But by inspection of equations 164, we can easily see that $(\partial F_i / \partial Y_i) = 0$ for all i, from i=1 to 2n. This immediately tells us that these systems are all statistically noncompressive. As discussed in section 3.4, this then implies that the exogenous source terms like $S^{[w]}$ are all zero. This implies that the entropy matrices all obey the usual sorts of linear dynamics familiar from quantum theory.

To follow the notation of section 3.4 exactly, in analyzing equation 162, we would use the notation "$a_1,...,a_{2n}$" to describe the operators which annihilate $Y_1,...,Y_{2n}$ respectively. But instead we will find it more convenient to use the notation "$a_1,...a_n,b_1,...,b_n$" to represent these operators.

Let us first consider the case where n=1 and f(X) is analytic, such that:

$$f(X) = \sum_{i=0}^{\infty} k_i X^i,$$

(166)

with Taylor series coefficients $k_i$. Following the same logic as in sections 3.4.2 and 3.4.3, we may immediately deduce that:

$$G_v = a^+b + b^+ \sum_{i=0}^{\infty} k_i a^i = a^+b + b^+ f(a) \tag{167}$$

Likewise, for the case where $n \geq 1$ and **f** is analytic, it is straightforward to generalize:

$$G_v = \sum_{i=0}^{\infty} \left( a_i^+ b_i + b_i^+ f_i(\underline{a}) \right) \tag{168}$$

The reification procedure of section 3.4.12, based on equation 160, then leads to:

$$G_z = \sum_{i=1}^{n} \left( (a_i^+ - a_i)(\frac{b_i + b_i^+}{2}) \right) + \sum_{i=1}^{n} \left( (b_i^+ - b_i) f_i(\frac{a+a^+}{2}) \right) \tag{169}$$

It is easy to see that $G_z$ here is antiHermitian. It is also easy to see that $G_z$ involves two sets of creation and annihilation operators, $\{a_i, a_i^+\}$ and $\{b_i, b_i^+\}$.

Within the general set of systems governed by equation 162, physicists are especially interested in the subset which can be represented as *Lagrangian* systems [Mandl]. In those cases, the ODE can be derived as the Lagrange-Euler equations for a Lagrangian:

$$L = \tfrac{1}{2}(\partial_t \underline{X})^2 - g(\underline{X}) \tag{170}$$

For such systems the Lagrange-Euler equations are:

$$\partial_t^2 X_i = -\frac{\partial g}{\partial X_i}(\underline{X}) \tag{171}$$

and the conserved energy or Hamiltonian is defined by:

$$H = \tfrac{1}{2}(\partial_t \underline{X})^2 + g(\underline{X}) \tag{172}$$

From equation 51, we may deduce, for any state $\underline{Y} = \underline{X} \oplus (\partial_t \underline{X})$, that:

$$H(\underline{Y}) = \underline{w}(\underline{Y})^T H_w \underline{w}(\underline{Y}) \tag{173}$$

where the operator $H_w$ may be

$$\tfrac{1}{2} \Sigma (b_i)^2 + g(\underline{a}) \tag{174}$$

or

$$\tfrac{1}{2} \Sigma b_i^+ b_i + g(\underline{a}) \tag{175}$$

or a variant which also replaces $g(\underline{a})$ with an equivalent symmetric operator $g'(\underline{a}^+, \underline{a})$.

After doing these calculations, is easy to see that none of these variations of $H_w$ will commute with $G_v$ or $G_w$. In fact, there is no reason why they should have to. For any operator $H_w$, $H_w$ is conserved if and only if:

$$0 = \partial_t(\underline{w}(\underline{Y})^T H_w \underline{w}(\underline{Y})) = \underline{w}(\underline{Y})^T (G_w^T H_w + H_w G_w) \underline{w}(\underline{Y}) \tag{176}$$

for all vectors $\underline{Y}$. This does *not* require that $G_w^T H_w + H_w G_w$ equal zero! It only requires that $G_w^T H_w + H_w G_w$ be a null operator as defined in section 3.3.7. Given any pair of operators $G_w$ and $H_w$, we may test whether this is in fact a null operator by using the methods given in that section.

### 3.5.3. Statistical Dynamics of Systems Governed By Equation 11

Systems governed by equation 11 are a special case of systems governed by equation 162. Therefore, they can be analyzed in either of two different ways: (1) exactly as a special case of equation 162, following all of the details already given in section 3.5.2; (2) as a unique case, using the special definitions given in equations 12-14, 25 and 26. This section will take the second approach. These special definitions lead to dynamical equations which look somewhat closer to the dynamical equations used in quantum field theory, which generally addresses similar special cases.

The first step in this analysis is to derive the dynamical equations for $\underline{X}^+$ and $\underline{X}^- \in C^n$ as defined in equations 12. Using only the definitions in equations 12, we may deduce:

$$\partial_t(X_k^+ + X_k^-) = \partial_t(X_k) = \dot{X}_k = im_k(X_k^+ - X_k^-) \tag{177}$$

Using both equations 12 and 11, we may deduce:

$$\partial_t(X_k^+ - X_k^-) = \partial_t(\frac{1}{im_k}\dot{X}_k) = \frac{1}{im_k}\left(-m_k^2(X_k^+ + X_k^-) + f_k(\underline{X}^+ + \underline{X}^-)\right) = im_k(X_k^+ + X_k^-) + \frac{1}{im_k}f_k(\underline{X}^+ + \underline{X}^-) \tag{178}$$

Adding and subtracting equations 177 and 178, we deduce:

$$\partial_t X_k^+ = im_k X_k^+ + \frac{1}{2im_k} f_k(\underline{X}^+ + \underline{X}^-) \tag{179}$$

$$\partial_t X_k^- = -im_k X_k^+ - \frac{1}{2im_k} f_k(\underline{X}^+ + \underline{X}^-) \tag{180}$$

If we had used the earlier definition of $\underline{y}$, from equation 10, then the methods of sections 3.4.2 and 3.4.3 immediately yield:

$$G_v \text{ (old definition )} = \sum_k \left( im_k(a_k^+ a_k - b_k^+ b_k) + \frac{1}{2im_k}(a_k^+ - b_k^+)f_k(\underline{a} + \underline{b}) \right) \tag{181}$$

The additional powers of m in the new definition of $\underline{y}$ (equation 14) have the effect of replacing every annihilation operator $a_i$ by $a_i/m_i^{1/2}$ and every creation operator $a_i^+$ by $a_i^+ m_i^{1/2}$, and likewise for the "b" operators. This gives us:

$$G_v \text{ (new definition )} = i\sum_k \left( m_k(a_k^+ a_k - b_k^+ b_k) - \frac{1}{2}\left(\frac{a_k^+ - b_k^+}{\sqrt{m_k}}\right)f_k\left(\left\{\frac{a_j + b_j}{\sqrt{m_j}}\right\}\right) \right) \tag{182}$$

Finally, in order to reify the statistical dynamics, we used the mixing transformation defined in equation 76. Using calculations parallel to those of section 3.4.12, we can deduce that this transformation has the effect:

$$a_k^+ \rightarrow a_k^+ - b_k \tag{183a}$$

$$b_k^+ \rightarrow b_k^+ - a_k \tag{183b}$$

$$a_k \rightarrow \tfrac{1}{2}(a_k + b_k^+) \tag{183c}$$

$$b_k \to \tfrac{1}{2}(b_k + a_k^+) \tag{183d}$$

Applying this to equation 182, we get:

$$G_z = i\sum_k m_k(a_k^+ a_k - b_k^+ b_k) - \tfrac{i}{2}\sum_k \left(\frac{a_k^+ - b_k - b_k^+ + a_k}{\sqrt{m_k}}\right) f_k\left(\left\{\frac{a_j + b_j^+ + b_j + a_j^+}{2\sqrt{m_j}}\right\}\right) \tag{184}$$

It is straightforward to show that $G_z$ is antiHermitian. (Note that we must exploit the fact that $a_k^+ - b_k - b_k^+ + a_k$ commutes with $a_j + b_j^+ + b_j + a_j^+$.) As in quantum field theory (QFT), the gain matrix is naturally expressed as "i" multiplied by an Hermitian operator. The biggest difference with QFT is that we use two sets of operators, "a" and "b," to represent the system. In addition, the low-order term contains a minus sign before $b_k^+ b_k$; this is absolutely unavoidable in a classical time-forwards calculation here, because the original ODE impose a symmetry between terms like exp(+iEt) and terms like exp(-iEt), their complex conjugates. The large fractions in the higher-order term in equation 184 are typical of the fractions used to define the ordinary "field operators" of QFT.

## 3.6 Statistical Dynamics of the Nonlinear Klein-Gordon Equation (15) with f($\varphi$) as $g\varphi^2$ or $g\varphi^3$

The purpose of this section is to derive the gain matrix $G_z$ for the nonlinear Klein-Gordon equation (15), for the special case where $\varphi$ is a real scalar and $f(\varphi)$ is chosen to be either $g\varphi^2$ or $g\varphi^3$, where g is a real scalar parameter.

Previous sections have already discussed dozens of equivalent ways to represent the probability $\Pr(\Phi)$ of the state $\Phi$ of a field. For second-order scalar PDE, the state $\Phi$ is defined by specifying both $\varphi(\underline{x})$ and $\partial_t \varphi(\underline{x})$, or equivalent information, across all points $\underline{x}$ in 3-dimensional space. Previous sections have described how the entropy matrix $S_w$ leads to a straightforward closure of turbulence, while the reified entropy matrix $S_z$ has striking similarities to the density matrix of QFT. The dynamics of $S_z$ are governed over time by the gain matrix $G_z$, as in equation 153 (with $S^{[z]}=0$ for second-order systems). Our goal here is to derive what $G_z$ is for this specific case.

For the case where $f(\varphi)=g\varphi^2$, the Fourier analysis sketched in equation 18 becomes:

$$\partial_t^2 \varphi(\underline{p}) = -|\underline{p}|^2 \varphi(\underline{p}) - M^2 \varphi(\underline{p}) + g\int \varphi(\underline{q})\varphi(\underline{p}-\underline{q}) d^3\underline{q} \tag{185}$$

Defining $\omega(\mathbf{p})$ as $\sqrt{|\mathbf{p}|^2 + M^2}$, as in equation 19, we can write this as:

$$\partial_t^2 \varphi(\underline{p}) = -\omega^2(\underline{p})\varphi(\underline{p}) + g\int \varphi(\underline{q})\varphi(\underline{p}-\underline{q}) d^3\underline{q} \tag{186}$$

Likewise, for the case where $f(\varphi)=g\varphi^3$, we have:

$$\partial_t^2 \varphi(\underline{p}) = -\omega^2(\underline{p})\varphi(\underline{p}) + g\int\int \varphi(\underline{q})\varphi(\underline{r})\varphi(\underline{p}-\underline{q}-\underline{r}) d^3\underline{q} d^3\underline{r} \tag{187}$$

Equations 186 and 187 may be viewed as extensions of the second-order ODE case, where instead of discrete subscripts (the "i" of $X_i$) we now have continuous subscripts, in effect (the $\mathbf{p}$ of $\varphi(\mathbf{p})$). Instead of vectors $\underline{X}$ over $R^n$, we are dealing with vectors over an ordinary Hilbert space. But we can still use the same analytical tools which we used for ordinary vectors in section 3.5. In particular, we have a choice of two valid representations for $\Phi$, one based on $\varphi$ and $\partial_t \varphi$, and the other based on $\varphi^+$ and $\varphi^-$, as in section 3.5.

For the first representation, it is more convenient here to define the two state variables as φ(**p**) and (∂$_t$φ(**p**))/ω(**p**). This new factor of ω(**p**) simplifies the representation of the system. Thus the operator a(**p**) will annihilate φ(**p**), but b(**p**) will annihilate (∂$_t$φ(**p**))/ω(**p**), *not* ∂$_t$φ(**p**).

Using these two sets of state variables, and applying the methods of sections 3.4.2 and 3.4.3 to equation 185, we get:

$$G_v = \int \mathbf{w}(\underline{p})\left(a^+(\underline{p})b(\underline{p}) - b^+(\underline{p})a(\underline{p})\right)d^3\underline{p} + \frac{g}{\mathbf{w}(\underline{p})}\iint b^+(\underline{p})a(\underline{p}-\underline{q})a(\underline{q})\,d^3\underline{p}\,d^3\underline{q}$$

(188)

Likewise, for the case f(φ)=g φ$^3$, we get:

$$G_v = \int \mathbf{w}(\underline{p})\left(a^+(\underline{p})b(\underline{p}) - b^+(\underline{p})a(\underline{p})\right)d^3\underline{p} + \frac{g}{\mathbf{w}(\underline{p})}\iiint b^+(\underline{p})a(\underline{p}-\underline{q}-\underline{r})a(\underline{q})a(\underline{r})\,d^3\underline{p}\,d^3\underline{q}\,d^3\underline{r}$$

(189)

As in section 3.5.2, we can reify these equations using the self-mixing similarity transformation. For the case f(φ)=g φ$^2$, we apply this to equation 188 to get:

$$G_z = \int \mathbf{w}(\underline{p})\left(a^+(\underline{p})b(\underline{p}) - b^+(\underline{p})a(\underline{p})\right)d^3\underline{p}$$
$$+ \frac{g}{4\mathbf{w}(\underline{p})}\iint (b^+(\underline{p}) - b(\underline{p})(a(\underline{p}-\underline{q}) + a^+(\underline{p}-\underline{q}))(a(\underline{q}) + a^+(\underline{q}))\,d^3\underline{p}\,d^3\underline{q}$$

(190)

Likewise, for the case f(φ)=g φ$^3$, we get:

$$G_z = \int \mathbf{w}(\underline{p})\left(a^+(\underline{p})b(\underline{p}) - b^+(\underline{p})a(\underline{p})\right)d^3\underline{p}$$
$$+ \frac{g}{8\mathbf{w}(\underline{p})}\iiint (b^+(\underline{p}) - b(\underline{p})(a(\underline{p}-\underline{q}-\underline{r}) + a^+(\underline{p}-\underline{q}-\underline{r}))(a(\underline{q}) + a^+(\underline{q}))(a(\underline{r}) + a^+(\underline{r}))\,d^3\underline{p}\,d^3\underline{q}\,d^3\underline{r}$$

(191)

These operators G$_z$ are elegant and antiHermitian, but they do not mirror some of the inelegant features of QFT.

For the second representation, we represent the state Φ at time t as {φ$^+$(**p**), φ$^-$(**p**)}, where φ$^+$ and φ$^-$ are defined exactly as in equations 12, except that m$_k$ is replaced by ω(**p**)$^{1/2}$. The analysis is an obvious extension of the analysis in section 3.5.3, and it yields the following extension of equation 182, for the case where f(φ)=g φ$^2$:

$$G_v = i\int \mathbf{w}(\underline{p})\left(a^+(\underline{p})a(\underline{p}) - b^+(\underline{p})b(\underline{p})\right)d^3\underline{p} - \frac{ig}{2}\iint \left(\frac{a^+(\underline{p}) - b^+(\underline{p})}{\sqrt{\mathbf{w}(\underline{p})}}\right)\left(\frac{a(\underline{p}-\underline{q}) + b(\underline{p}-\underline{q})}{\sqrt{\mathbf{w}(\underline{p}-\underline{q})}}\right)\left(\frac{a(\underline{q}) + b(\underline{q})}{\sqrt{\mathbf{w}(\underline{q})}}\right)d^3\underline{p}\,d^3\underline{q}$$

(192)

which may be written more compactly as:

$$G_v = iH_0 + iH_I,$$ (193)

where:

$$H_0 = \int \mathbf{w}(\underline{p})\left(a^+(\underline{p})a(\underline{p}) - b^+(\underline{p})b(\underline{p})\right)d^3\underline{p}$$

(194)

and:

$$H_I = -\frac{g}{2}\iint \left(\frac{a^+(\underline{p})-b^+(\underline{p})}{\sqrt{\mathbf{w}(\underline{p})}}\right)\left(\frac{a(\underline{p}-\underline{q})+b(\underline{p}-\underline{q})}{\sqrt{\mathbf{w}(\underline{p}-\underline{q})}}\right)\left(\frac{a(\underline{q})+b(\underline{q})}{\sqrt{\mathbf{w}(\underline{q})}}\right) d^3\underline{p}\, d^3\underline{q}$$

(195)

Likewise, for the case $f(\varphi)=g\varphi^3$, we still obtain equations 193 and 194 but we replace the expression for $H_I$ by:

$$H_I = -\frac{g}{2}\iiint \left(\frac{a^+(\underline{p})-b^+(\underline{p})}{\sqrt{\mathbf{w}(\underline{p})}}\right)\left(\frac{a(\underline{p}-\underline{q}-\underline{r})+b(\underline{p}-\underline{q}-\underline{r})}{\sqrt{\mathbf{w}(\underline{p}-\underline{q}-\underline{r})}}\right)\left(\frac{a(\underline{q})+b(\underline{q})}{\sqrt{\mathbf{w}(\underline{q})}}\right)\left(\frac{a(\underline{r})+b(\underline{r})}{\sqrt{\mathbf{w}(\underline{r})}}\right) d^3\underline{p}\, d^3\underline{q}\, d^3\underline{r}$$

(196)

In order to reify these systems, we use the mixing transformation used in section 3.5.3, generalized to the PDE case (see [Werbos93],p.319), which replaces equation 183 by:

$$a^+(\mathbf{p}) \rightarrow a^+(\mathbf{p}) - b(-\mathbf{p}) \tag{197a}$$

$$b^+(\mathbf{p}) \rightarrow b^+(\mathbf{p}) - a(-\mathbf{p}) \tag{197b}$$

$$a(\mathbf{p}) \rightarrow \tfrac{1}{2}(a(\mathbf{p}) + b^+(-\mathbf{p})) \tag{197c}$$

$$b(\mathbf{p}) \rightarrow \tfrac{1}{2}(b(\mathbf{p}) + a^+(-\mathbf{p})) \tag{197d}$$

Applying this to equations 193-195, for the case $f(\varphi)=g\varphi^2$, we derive:

$$G_z = iH_0 - iH_I', \tag{198}$$

where:

$$H_I' = \frac{g}{8}\iint \left(\frac{a^+(\underline{p})-b(-\underline{p})-b^+(\underline{p})+a(-\underline{p})}{\sqrt{\mathbf{w}(\underline{p})}}\right)\left(\frac{a(\underline{p}-\underline{q})+b^+(\underline{q}-\underline{p})+b(\underline{p}-\underline{q})+a^+(\underline{q}-\underline{p})}{\sqrt{\mathbf{w}(\underline{p}-\underline{q})}}\right)\left(\frac{a(\underline{q})+b^+(-\underline{q})+b(\underline{q})+a^+(-\underline{q})}{\sqrt{\mathbf{w}(\underline{q})}}\right) d^3\underline{p}\, d^3\underline{q}$$

(199)

Equation 199 may be expressed more compactly as:

$$H_I' = g\iint (A(\underline{p}) - B(\underline{p}))(A(\underline{q}-\underline{p}) + B(\underline{q}-\underline{p}))(A(-\underline{q}) + B(-\underline{q})) d^3\underline{p}\, d^3\underline{q} \;,$$

(200)

where we define:

$$A(\underline{p}) = \frac{a^+(\underline{p}) + a(-\underline{p})}{2\sqrt{\mathbf{w}(\underline{p})}}$$

(201a)

$$B(\underline{p}) = \frac{b^+(\underline{p}) + b(-\underline{p})}{2\sqrt{\mathbf{w}(\underline{p})}}$$

(201b)

Note that equations 201 are more or less identical to the usual definitions of field operators used in QFT [Mandl]. (Also, they are different from the A and B defined in [Werbos93], which were used as tools to prove that $H_I'$ is Hermitian.) Likewise, for the case $f(\varphi)=\varphi^3$, the similarity transformation in equation 197 yields:

$$H_I' = g \iiint (A(\underline{p}) - B(\underline{p}))(A(\underline{q}+\underline{r}-\underline{p}) + B(\underline{q}+\underline{r}-\underline{p}))(A(-\underline{q}) + B(-\underline{q}))(A(-\underline{r}) + B(-\underline{r})) d^3\underline{p}\, d^3\underline{q}\, d^3\underline{r} \quad,$$
(202)

The similarity to QFT is very striking, and we find it extremely hard to believe that these parallels are all just a coincidence. Nevertheless, we still require *two* sets of field operators, A and B, instead of just one, in order to represent a real scalar field theory.

## 3.7. Conclusions From Section 3

This section has addressed the general problem of how to map a probability distribution, Pr(**X**) or Pr(Φ), for the state of a dynamical system at any time, into a more convenient representation over Fock space. Some of these representations (like the entropy matrix $S_w$) permit a straightforward "closure of turbulence" in the general case. Other representations focused on the special types of system handled well by QFT result in dynamics extremely close to the dynamics of density matrices in QFT. Both types of representation also provide a kind of measurement formalism, a way to calculate field expectation values by means of a trace formula essentially the same as the formula used in QFT. But here the trace formulas can be derived analytically rather than assumed apriori.

In the end, however, there are still a few differences between the gain matrices here and the iH matrices normally seen in QFT. Section 4 will argue that there is an exact equivalence, for some PDE; however, it may not be possible to see this equivalence when we limit ourselves to mappings FROM information at time t TO information at time t. Consider, for example, the old trick of representing a positron moving forward from creation at time t to absorption at time τ as a hole moving backwards from time τ to time t; the underlying equivalence can only be understood *across* time. (Certainly this is relevant to the sign of $b^+b$ in $H_0$ above.) As another example, the famous equivalence between the sine-Gordon system (a special case of equation 15!) and the massive Thirring model could be proven only by considering the resulting Feynman integrals across time [Coleman].

We would conjecture that the dynamic systems in Fock space derived above are exactly equivalent to the usual scalar Klein-Gordon field theories (at least for $f=g\,\varphi^2$) if one considers the Feynman integrals over space time, and allows for the fact that our formulation here allows for a more general boundary conditions than the usual QFT version. Section 4 will explain the basis for this conjecture. In general, however, we hope that mathematicians more skilled than us will take up the critical task of moving these topics from the realm of engineering and conjecture (however well justified) to the realm of theorems and proofs.

## 4. Effective Correlated Sources (ECS): The Bridge Between Statistics and Quantum Theory

### 4.1. Goals and Conclusions

The main goal of this section is to show that the predictions of the Backwards Time Interpretation (BTI) of quantum mechanics correspond exactly to the predictions of ordinary quantum field theory (2QFT), for a tiny model field theory, the real scalar $\varphi^3$ field theory, which is based on equation 15 with $f(\varphi)=g\,\varphi^2$. (The mathematician should note that this is called "$\varphi^3$" theory rather than "$\varphi^2$" theory, based on the Lagrangian for the system.) Another goal is to demonstrate tools which could be used in the analysis of other field theories as well. Such an equivalence does not hold for real scalar $\varphi^4$ theory; however, the standard models of physics are essentially cubic (in their Lagrangians) rather than quartic.

Section 2 described how important this is to physics. Section 2.6 described how to implement BTI by combining: (1) a measurement formalism, justified by the trace expectation value formulas derived in section 3; (2) methods for predicting the statistical moments coming out of an experiment, in some representation, as a function of what goes into the experiment. This section will focus on the second of these issues, the only area where basic loose ends still exist after the discussions of sections 2 and 3.

Section 3 derived the laws of evolution for the statistical moments of a dynamical system, in the general case, for about 20 representations of those moments, without regard to the special needs of quantum theory. Section 3.6 showed that the laws of evolution for a major class of systems are in fact true Schrodinger equations, based on field operators which could have been taken from a standard textbook on 2QFT. But the Schrodinger equations for the real scalar $\varphi^3$ and $\varphi^4$ field theories had certain differences from the usual Schrodinger equations for those field theories, as seen in texts like [Mandl,Weinberg]. The biggest difference was that the Schrodinger equations derived from statistics require a *pair* of field operators, A(**p**) and B(**p**), rather than the single field operator $\varphi$(**p**) used in 2QFT.

In principle, this gap need not be worrisome. The statistical approach still gives us a wide variety of choices in PDE and Schrodinger equations. There is no reason why the true map from PDE to Schrodinger equations has to correspond to the map used in 2QFT; it is good enough to have access to a wide class of Schrodinger equations. Nevertheless, for the case of $\varphi^3$, we do not have to learn to live with this kind of gap. The statistical approach and 2QFT map to the same predictions, in the end.

Section 4.2 will explain the reason behind the apparent gap, using a simple example of a second-order system to illustrate the underlying problem. In essence, the approach of section 3 requires an extra pair of operators (b, b$^+$) because it allows for *arbitrary* initial conditions at an initial time $t_-$. The arbitrary initial conditions require the use of a propagator or Green's function which is a two-by-one matrix rather than a scalar; this is what doubles the apparent complexity. Quantum field theory, on the other hand, implicitly assumes that the initial conditions can be represented in a different way, as *effective sources* of incoming particles. ([Schwinger] makes this quite explicit.) When we represent the initial conditions in this way, we can compute the statistics required by BTI by using only a *scalar* propagator. That, in turn, makes it possible to formulate the statistics of the same scalar PDE using only a single pair of operators. Section 4.3 will show how this can be done for the case of $\varphi^3$ and $\varphi^4$ field theories.

## 4.2. Initial Conditions Versus ECS: A Simple Example

Consider the following simple dynamical system, a special case of equation 11:

$$\partial_t^2 X = -m^2 X + cX^2 \tag{203}$$

The state of this system at any time t may be specified as a 2-component vector, **Y**, defined as $X \oplus (\partial_t X)$ or as $X_+ \oplus X_-$, as discussed in section 3.5. We can use the methods of section 3.5 to derive how the various representations of Pr(**Y**) evolve over time. Those methods tell us how to start from an arbitrary probability distribution Pr(**Y**($t_-$)), at some time $t_-$, in order to predict probabilities or correlations or expectation values at later times t.

Section 3 used differential equations to describe the evolution of probabilities. Here we will use integral equations, which are perhaps more common in practice in modern high-energy physics.

Equation 203 may be written equivalently as:

$$X(t) = G_1(t - t_-)X(t_-) + G_2(t - t_-)(\partial_t X(t_-)) + \int_{t_-}^{t} G_2(t - \boldsymbol{t})S(\boldsymbol{t})d\boldsymbol{t}, \tag{204}$$

where:

$$G_1(\tau) = \cos m\tau \tag{205a}$$

$$G_2(\tau) = (1/m) \sin m\tau \tag{205b}$$

$$S(\tau) = cX^2(\tau) \tag{205c}$$

It is obvious why we need two propagators here ($G_1$ and $G_2$) in order to cope with general initial conditions at time $t_-$! This is the basis for our conclusions in section 4.1. However, suppose that we could find a function s(t), for t<$t_-$, such that:

$$X(t_-) = \int_{-\infty}^{t_-} G_2(t_- - \boldsymbol{t}) s(\boldsymbol{t}) d\boldsymbol{t}$$

and: (206a)

$$\partial_t X(t_-) = \partial_t \bigg|_{t=t_-} \int_{-\infty}^{t_-} G_2(t - \boldsymbol{t}) s(\boldsymbol{t}) d\boldsymbol{t}$$

(206b)

In this representation, we can replace equation 204 by:

$$X(t) = \int_{-\infty}^{t} G_2(t - \boldsymbol{t}) S(\boldsymbol{t}) d\boldsymbol{t}$$

(207)

where $G_2$ is as before and:

$$S(\tau) = cX^2(\tau) \quad \tau \geq t_- \quad (208a)$$

$$S(\tau) = s(\tau) \quad \tau < t \quad (208b)$$

We would describe equations 206 as an *effective source* representation of the initial conditions. Schwinger's original development of source theory [Schwinger] – the foundation of modern functional integral quantum field theory (FIQFT) [Zinn-Justin] – relies on exactly this approach to characterizing the particles incoming to a scattering experiment. Once we assume such a representation, equation 207 allows us to get by, using only a single scalar propagator.

The situation is essentially the same for *any* second-order system, ODE or PDE. Thus the effective source representation is fundamental to the compact, QFT-like statistical analysis of systems like the nonlinear Klein-Gordon equation. One might imagine that things will become different when we consider first-order PDE like the Dirac equation; however, our intuition suggests strongly that the tools needed to explain fermionic statistics (hinted at in section 2) will again create a need to use the ECS representation, for a compact formulation of the field theory.

Our discussion so far has been deterministic in nature. The next section will address the full stochastic case. In essence, it will simply take expectation values, as we did in section 3 when we defined **u**=<**u**(**X**)>. Thus the generalization of equation 207 will allow us to map from the statistics of the source to the statistics of the field at later times. Because our initial conditions will be statistical in nature, accounting for cross-correlations, we will call them *effective correlated sources* (ECS).

The ECS approach will lead naturally to the usual sorts of Feynman integrals, which in turn make it easy to perform Dirac-like flips in representation for cosmetic purposes, as in ordinary QFT.

## 4.3. Application of ECS to $\boldsymbol{\varphi}^3$ and Then $\boldsymbol{\varphi}^4$ Field Theories

This section will show how to calculate the statistical moments coming out of a scattering experiment, as a function of initial conditions expressed as effective correlated sources. In order to avoid unnecessary complexity, we will formulate the calculations on a true 3+1 dimensional basis; this can replicate the usual Feynman integrals of QFT, even though it does not make use of a Schrodinger equation as such. (The same is true for FIQFT! [Zinn-Justin].) This will require substantial changes in our notation from section 3.

Let us first consider the case of a nonlinear Klein-Gordon equation with $f(\varphi)=g\varphi^2$. In parallel with equations 207 and 208, we may express this system as:

$$\boldsymbol{\varphi}(g, x) = \int G(x-y) S(y) d^4 y = \int G(x-y)(s(y) + g\boldsymbol{\varphi}^2(g, y)) d^4 y \quad ,$$

(209)

where G is the usual retarded propagator for the Klein-Gordon equation, where x and y are vectors in the usual 3+1-D Minkowski space *M*, and where the integrals are taken over all space-time. Notice that we have chosen to include the coupling coefficient g itself as a parameter of φ, for use in later calculations. In other words, we will find it convenient to consider the entire *family* of solutions φ(x) which would result from the initial conditions s(y) across different values of the coupling coefficient g.

For any real scalar field $\varphi(x)$, we will now define the state $\Phi$ of the field to be $\{\varphi(x), \text{ all } x \in M\}$ – the values of the field across *all* points in space-time. We will define $\underline{u}(\Phi)$ as a vector in the Fock space $F(M \to R^1)$ made up of components

$$u_0 = 1$$
$$u_1(x) = \varphi(x) \quad (= \varphi(\underline{x}, t_x))$$
$$u_2(x, y) = \varphi(x)\varphi(y) \quad (= \varphi(\underline{x}, t_x)\varphi(\underline{y}, t_y))$$
$$...$$

(210)

Note that $\underline{u}$ includes all correlations across space-time, not just correlations across space at a particular time.

Let us define:

$$\underline{u}(g) = <\underline{u}(\{\varphi(g,x), \text{ all } x\})> \tag{211}$$

Note that $\underline{u}(0)$ is simply the statistical moments of the function $\varphi(0,x)$ which, by equation 209, equals:

$$\boldsymbol{j}(0,x) = \int G(x-y)s(y)d^4y \tag{212}$$

We would call $\varphi(0,x)$ the *linear wave* generated by the source terms. In practical terms, $\varphi(0,x)$ is what really represents our initial conditions. Even in Schwinger's use of source theory [Schwinger], he actually "picks out" terms in the global vacuum probability amplitude expression which correspond to different modes of $\varphi(0,x)$; in other words, he derives probability amplitudes by using $\varphi(0,x)$ as the actual effective initial conditions. Here we will be somewhat more direct about this. Our initial goal is to derive the mapping from initial conditions $\varphi(0,x)$ to the outgoing *nonlinear wave* $\varphi(g, x)$, by deriving the matrix $W_u$ which maps:

$$\underline{u}(g) = W_u \underline{u}(0) \tag{213}$$

Note that the matrix $W_u$ plays a role similar to that of the S matrix in 2QFT. Later, when we scale and reify the statistical moments, we will derive a new matrix $W_z$ which plays a role essentially identical to that of the usual S matrix.

It is often possible to derive $W_u$ by performing a Taylor series expansion of $\underline{u}(g)$ as a function of g about g=0:

$$\underline{u}(g) = \sum_{k=0}^{\infty} \frac{g^k}{k!} \partial_g^k \bigg|_{g=0} \underline{u}(g)$$

(214)

We do not claim that this kind of Taylor series will always converge, for all field theories and all coupling coefficients g; indeed, our goal in this subsection is to match up to the usual Feynman integrals of 2QFT and perturbation theory, which themselves do not always converge.

As in section 3.6 and in QFT, we will find it most convenient to work in Fourier transforms here. Under a full 3+1-D Fourier transform, equation 209 becomes:

$$\boldsymbol{j}(g,p) = \frac{1}{\Delta(p)} \left( s(p) + g \int \boldsymbol{j}(g, p-q)\boldsymbol{j}(g,q)d^4q \right) ,$$

where: (215)

$$\Delta(p) = p^2 + M^2 \quad (= \underline{p}^2 + M^2 - p_0^2) \tag{216}$$

In order to compute the derivatives required in equation 214, let us first see what happens when we differentiate equation 215 with respect to g:

$$\partial_g \boldsymbol{j}(g,p) = \frac{1}{\Delta(p)} \left( \int \boldsymbol{j}(g,p-q)\boldsymbol{j}(g,q) d^4q + g\partial_g \int \boldsymbol{j}(g,p-q)\boldsymbol{j}(g,q) d^4q \right)$$

(217)

More generally, we claim that:

$$\partial_g^n \boldsymbol{j}(g,p) = \frac{1}{\Delta(p)} \left( n\partial_g^{n-1} \int \boldsymbol{j}(g,p-q)\boldsymbol{j}(g,q) d^4q + g\partial_g^n \int \boldsymbol{j}(g,p-q)\boldsymbol{j}(g,q) d^4q \right)$$

(218)

Equation 218 can be verified easily by mathematical induction. For the case n=1, it is equivalent to equation 217. For the case n+1, it may be derived by simply differentiating the equation for case n with respect to g. Equation 218 implies:

$$\partial_g^n \Big|_{g=0} \boldsymbol{j}(g,p) = \frac{n}{\Delta(p)} \partial_g^{n-1}\Big|_{g=0} \int \boldsymbol{j}(g,p-q)\boldsymbol{j}(g,q) d^4q$$

(219)

Following the approach of section 3.4.2, we may deduce:

$$\partial_g^n \Big|_{g=0} \underline{u}(g,p) = n G_u^{[4]} \partial_g^{n-1}\Big|_{g=0} \underline{u}(g,p)$$

(220)

where we define:

$$G_u^{[4]} = \iint \frac{N(p)c^+(p)}{\Delta(p)} c(p-q)c(q) d^4p\, d^4q$$

(221)

where we use "clean" creation and annihilation operators $c^+(p)$ and $c(p)$ like the clean operators of section 3.4.2. Here, however, the operators operate "in four dimensions," to create or annihilate modes across *space-time*, not just space! By iterating over equation 220, we easily deduce:

$$\partial_g^n \Big|_{g=0} \underline{u}(g,p) = n!(G_u^{[4]})^n \underline{u}(0,p)$$

(222)

Substituting back into equations 214 and 213, we then deduce:

$$W_u = \sum_{k=0}^{\infty} (gG_u^{[4]})^k$$

(223)

This matrix $W_u$ already has strong similarities to the usual S matrix. When the usual contraction procedures [Mandl, ch.13] are used to work out the value of $W_u$, they yield Feynman integrals with the usual energy denominators $\Delta^{-1}(p)$. The commutators of our *four-dimensional* creation and annihilation operators directly yield the usual four-dimensional Dirac delta functions, functions which are derived in a far more cumbersome way in 2QFT as the result of integration over kludgey oscillation terms [Mandl, equation 15.11].

Nevertheless, in order to avoid the problem of "infinite regress" or "closure of turbulence" discussed at length in section 3, we must again scale and reify the statistical moments. If our only goal were to perform the calculations correctly and efficiently, we should consider using representations like the entropy matrix $S_w$. However, in order to establish correspondence with 2QFT, we will limit ourselves for now to the **y** and **z** representations.

In parallel with equation 19, we will define:

$$v_n(p_1,\ldots,p_n) = u_n(p_1,\ldots,p_n)\sqrt{\frac{\Delta(p_1)\Delta(p_2)\ldots\Delta(p_n)}{n_1!n_2!\ldots n_a!}} \quad,$$

(224)

where $\alpha$ is the number of distinct values for p in the sequence $p_1,\ldots,p_n$, and where $n_i$ represents the number of occurrences of the i-th distinct value in that sequence. Since **v** is really just a linear similarity transformation of **u** (based on a diagonal transformation here!), we may easily deduce:

$$\underline{v}(g) = W_v \, \underline{v}(0),$$

(225)

where:

$$W_v = \sum_{k=0}^{\infty} (gG_v^{[4]})^k \quad,$$

(226)

and, as in section 3.4.3, we may derive $G_v^{[4]}$ from a simple multiplication of $G_u^{[4]}$ on the left and right, which yields:

$$G_v^{[4]} = \iint \left(\frac{a^+(p)}{\sqrt{\Delta(p)}}\right)\left(\frac{a(p-q)}{\sqrt{\Delta(p-q)}}\right)\left(\frac{a(q)}{\sqrt{\Delta(q)}}\right) d^4p\, d^4q \quad,$$

(227)

where $a^+(p)$ and $a(p)$ are "physical" creation and annihilation operators over four dimensions.

In order to reify the statistics, we first define new elementary reification operators analogous to those of section 3.3.4:

$$S_+(c) = \exp\left(c\int a^+(p)a^+(-p)d^4p\right)$$

(228a)

$$S_-(c) = \exp\left(c\int a(p)a(-p)d^4p\right)$$

(228b)

By using commutator algebra similar to that used in sections 3.4.4 and 3.4.12, which led up to equations 160, we may derive the effects of the corresponding similarity transformations here:

$$S_+(c): \quad a^+(p) \to a^+(p); \; a(p) \to a(p) - 2ca^+(-p) \quad (229a)$$

$$S_-(c): \quad a(p) \to a(p); \; a^+(p) \to a^+(p) + 2ca(-p) \quad (229b)$$

Using these operators, we may reify the statistics by defining:

$$\underline{z} = S_+(+\tfrac{1}{2})S_-(\tfrac{1}{4})\underline{v} \tag{230}$$

This gives us:

$$\underline{z}(g) = W_z \, \underline{z}(0), \tag{231}$$

where:

$$W_z = \sum_{k=0}^{\infty} (gG_z^{[4]})^k \quad,$$

(232)

and:

$$G_z^{[4]} = \iint \frac{(a^+(p)+a(-p))(a(p-q)-a^+(q-p))(a(q)-a^+(-q))}{2\sqrt{\Delta(p)\Delta(p-q)\Delta(q)}} d^4p\, d^4q$$

(233)

When this matrix $G_z^{[4]}$ matrix is used to make W-matrix calculations,
the triple-creation and triple-annihilation terms which seem to be present in the numerator actually drop out, exactly as in 2QFT. Thus we have the same exact energy denominators as in the $H_I$ matrix used the same way in 2QFT calculations in the interaction picture, and the effective numerator has the same form, $k(aaa^+ + aa^+a^+)$. Aside from scalar factors which are easily adjusted to tighten up every last detail, the results are exactly the same Feynman integrals (in the $W_z$ matrix calculation) as in the usual S matrix calculations of 2QFT. We believe that the propagator is converted from a retarded propagator to a symmetric propagator by the usual preliminary stages of calculation, exactly as in S matrix calculation [Mandl].

When this same procedure is used on the $\varphi^4$ field theory, however, equation 233 is replaced by:

$$G_z^{[4]} = \iint \frac{(a^+(p)+a(-p))(a(p-q-r)-a^+(q-p-r))(a(q)-a^+(-q))(a(r)-a^+(-r))}{2\sqrt{\Delta(p)\Delta(p-q)\Delta(q)\Delta(r)}} d^4p\, d^4q\, d^4r$$

(234)

We have not taken the trouble to prove that this is not equivalent to the 2QFT version of $\varphi^4$, or to prove that equivalence is impossible under other reification approaches; however, such an equivalence seems unlikely to us, based on our sense of how the Lie group generated by $S_+$ and $S_-$ appears to work.

In summary, we have found an exact equivalence between the statistics generated by the "classical" nonlinear Klein-Gordon equation in the $\varphi^3$ case and the predictions of the usual 2QFT which is normally associated with that equation. The W matrix and the S matrix match. This, combined with the larger context discussed in section 2, establishes an equivalence between the BTI view of quantum theory and the conventional views, for this common toy field theory. But for another common toy field theory, $\varphi^4$, we have not found such an equivalence.